\documentclass[12pt]{article}
\usepackage{amsmath,amssymb,amsthm,amsxtra,overpic,bbm,bm,epsfig,sub
figure}
\usepackage{mathrsfs}
\usepackage{graphicx}
\usepackage{color}
\usepackage{comment}
\usepackage{epstopdf}
\usepackage{float}
\usepackage{slashed,stmaryrd}

\def\thefootnote{\fnsymbol{footnote}}

\begin{document}

\vspace{0.2cm}

\begin{center}
{\Large\bf Renormalization-group evolution of the fermion mass matrices}
\end{center}

\vspace{0.2cm}

\begin{center}
{\bf Zhi-cheng Liu$^{1, 2}$} \footnote{E-mail:liuzhicheng@ihep.ac.cn}
\\
{\small $^{1}$Institute of High Energy Physics, Chinese Academy of
Sciences, Beijing 100049, China \\
$^{2}$School of Physical Sciences, University of Chinese Academy of
Sciences,
Beijing 100049, China } \\
\end{center}

\vspace{1.5cm}
\begin{abstract}
The one-loop renormalization-group equations (RGEs) running behavior of quark and lepton mass matrices with general structures are studied simultaneously.
Suppose the non-linear terms of RGEs are dominated by the Yukawa couplings of top quark and $\tau$ lepton, 
the unitary matrices that diagonalize the mass matrices of up-type quark and charged lepton in Hermitian basis are found as invariant with energy scale.
Based on this result, we can decouple the RGEs and obtain the RGE corrected mass matrices of fermion.
As examples, we consider the renormalization-group evolution of fermion matrices with four or five texture zeros and of lepton mass matrices which realize the tribimaximal mixing.
\end{abstract}
\begin{flushleft}
\hspace{0.8cm} PACS number(s): 14.60.Pq, 14.65.−q
\end{flushleft}

\def\thefootnote{\arabic{footnote}}
\setcounter{footnote}{0}

\newpage

\section{Introduction}

Flavor is one of the most important open questions in particle physics. In the standard model (SM), the Higgs mechanism that generate masses for gauge bosons and charged fermions
can not predict the values of fermion masses and mixing angles. To seek the dynamics of mass generation, one should appeal to new physics beyond the SM which is expected to appear at high scale. Phenomenologically, a less ambitious goal is to devise some predictable textures of fermion mass matrices from experimental data. Then the mass textures might provide some hints of the flavor symmetry or new mass generation.

When the proposed mass textures are confronted with the experimental data at low energies, it is worthwhile to explore how the relevant parameters change via RGEs.
There are two equivalent ways to consider the RGE running effects. The first one is extracting the masses and mixing angles from mass matrices with some special structures at a high scale, 
and then running these physical quantities down to the low scale via the relevant RGEs; another way is evolving the mass matrices directly, and then calculating the masses and mixing angles
from the modified mass matrices at the low scale. Here we adopt the second approach since it has the advantage of showing how the original mass matrices are modified by RGE running effects.
But one has to diagonalize a more complicated mass matrices to obtain the physical quantities if the hypothetical structures are not stable against RGE running. 
If a specific mass texture that is compatible with all experimental data at low energy scale are unchanged after RGE running effect are taken in account, it might be a candidate of 
the true pattern of fermion mass matrices at hight scale.
Refs. \cite{Albright:1988im,Grzadkowski:1987tf,Grzadkowski:1987wr,Giudice:1992an,Xing:2015sva} have considered how the textures of fermion mass matrices are spoiled by the RGE running effect
in a approximative way.  
In this work, we find a more exact method and consider the renormalization-group evolution of quark and lepton mass matrices with general structures simultaneously, since their RGEs are entangled with each other. The result might be helpful in understanding flavor physics beyond the standard electroweak model.

The rest of this paper is organized as follows, In section 2, we universally derive the RGE-corrected mass matrices of fermion, where both Dirac and Majorana neutrinos are considered.
Section 3 and 4 is devoted to some particular examples. Finally, we summarize our main results in section 5.

\section{Renormalization-group evol{\tiny }ution of the fermion mass matrices}
\subsection{The case of Dirac neutrinos}
After spontaneous breaking of electroweak symmetry, the fermion mass terms in SM and MSSM are given by
\begin{eqnarray}
\mathcal{L}^{}_{\rm SM} &\supset& \frac{v}{\sqrt 2} \bar u^{}_{L} Y^{}_{\rm u} u^{}_{R} + \frac{v}{\sqrt 2} \bar d^{}_{L} Y^{}_{\rm u} d^{}_{R}
+ \frac{v}{\sqrt 2} \bar \nu^{}_{L} Y^{}_{\nu} \nu^{}_{R} + \frac{v}{\sqrt 2} \bar l^{}_{L} Y^{}_{l} l^{}_{R} +  {\rm h.c.}\;, \nonumber \\
\mathcal{L}^{}_{\rm MSSM} &\supset& \frac{v \sin \beta }{\sqrt 2} \bar u^{}_{L} Y^{}_{\rm u} u^{}_{R} + \frac{v \cos \beta }{\sqrt 2} \bar d^{}_{L} Y^{}_{\rm u} d^{}_{R}
+ \frac{v \sin \beta }{\sqrt 2} \bar \nu^{}_{L} Y^{}_{\nu} \nu^{}_{R} + \frac{v \cos \beta }{\sqrt 2} \bar l^{}_{L} Y^{}_{l} l^{}_{R} +  {\rm h.c.}\;,
\end{eqnarray}
where $Y^{}_{\rm f}$ with ${\rm f=u,d},\nu,l$ are Yukawa matrices, $\tan \beta$ is the ratio of Higgs vacuum expectation values and $v=246$ GeV.
The differential form of the one-loop renormalization-group equations of Dirac fermion mass matrices are given by \cite{Lindner:2005as, Xing:2011zza}
\begin{eqnarray}
16 \pi^2 \frac{{\rm d} M^{}_{\rm u}}{{\rm d}t} &=& \left[ \alpha^{}_{\rm u} + \frac{3}{2} \left( b Y^{}_{\rm u} Y^{\dagger}_{\rm u} + c Y^{}_{\rm d} Y^{\dagger}_{\rm d} \right) \right]
M^{}_{\rm u} \;, \nonumber \\
16 \pi^2 \frac{{\rm d} M^{}_{\rm d}}{{\rm d}t} &=& \left[ \alpha^{}_{\rm d} + \frac{3}{2} \left( c Y^{}_{\rm u} Y^{\dagger}_{\rm u} + b Y^{}_{\rm d} Y^{\dagger}_{\rm d} \right) \right]
M^{}_{\rm d} \;, \nonumber \\
16 \pi^2 \frac{{\rm d} M^{}_{{\nu}}}{{\rm d}t} &=& \left[ \alpha^{}_{\nu} + \frac{3}{2} \left( b Y^{}_{\nu } Y^{\dagger}_{\nu } + c Y^{}_{l} Y^{\dagger}_{l} \right) \right]
M^{}_{\nu} \;, \nonumber \\
16 \pi^2 \frac{{\rm d} M^{}_{{l}}}{{\rm d}t} &=& \left[ \alpha^{}_{l} + \frac{3}{2} \left( c Y^{}_{\nu } Y^{\dagger}_{\nu } + b Y^{}_{l} Y^{\dagger}_{l} \right) \right]
M^{}_{l} \;. \label{1}
\end{eqnarray}
where $t\equiv \ln \left( \mu/\Lambda^{}_{\rm EW} \right)$ with $\mu$ being the renormalization energy scale and $\Lambda^{}_{\rm EW}$ being the electroweak energy scale,
$M^{}_{\rm f}$ with ${\rm f=u,d},\nu,l$ are mass matrices.
In the framework of the SM
\begin{eqnarray}
&& b=1 \;,\qquad c = -1 \;,\nonumber \\
&& \alpha^{}_{\rm u} = - \frac{17}{20} g^2_1 - \frac{9}{4} g^2_2 - 8 g^2_3 + R^{}_{\rm SM} \;, \nonumber \\
&& \alpha^{}_{\rm d} = - \frac{1}{4} g^2_1 - \frac{9}{4} g^2_2 - 8 g^2_3 + R^{}_{\rm SM} \;, \nonumber \\
&& \alpha^{}_{\nu} = - \frac{9}{20} g^2_1 - \frac{9}{4} g^2_2 + R^{}_{\rm SM} \;, \nonumber \\
&& \alpha^{}_{l} = - \frac{9}{4} g^2_1 - \frac{9}{4} g^2_2 + R^{}_{\rm SM} \;, \label{2}
\end{eqnarray}
with $R^{}_{\rm SM} = {\rm Tr} \left( 3 Y^{}_{\rm u} Y^{\dagger}_{\rm u} + 3 Y^{}_{\rm d} Y^{\dagger}_{\rm d} + Y^{}_{\nu} Y^{\dagger}_{\nu} + Y^{}_{l} Y^{\dagger}_{l} \right) $.
In framework of the minimal supersymmetry standard model (MSSM)
\begin{eqnarray}
&& b= 2 \;,\qquad c = \frac{2}{3} \;,\nonumber \\ 
&& \alpha^{}_{\rm u} = - \frac{13}{15} g^2_1 - 3 g^2_2 - \frac{16}{3} g^2_2 + {\rm Tr} \left( 3 Y^{}_{\rm u} Y^{\dagger}_{\rm u} + Y^{}_{\nu} Y^{\dagger}_{\nu} \right) \;, \nonumber \\
&& \alpha^{}_{\rm d} = - \frac{7}{15} g^2_1 - 3 g^2_2 - \frac{16}{3} g^2_2 +  {\rm Tr} \left( 3 Y^{}_{\rm d} Y^{\dagger}_{\rm d} + Y^{}_{l} Y^{\dagger}_{l} \right) \;, \nonumber \\
&& \alpha^{}_{\nu} = - \frac{3}{5} g^2_1 - 3 g^2_2 + {\rm Tr} \left( 3 Y^{}_{\rm u} Y^{\dagger}_{\rm u} + Y^{}_{\nu} Y^{\dagger}_{\nu} \right) \;, \nonumber \\
&& \alpha^{}_{l} = - \frac{9}{5} g^2_1 - 3 g^2_2 + {\rm Tr} \left( 3 Y^{}_{\rm d} Y^{\dagger}_{\rm d} + Y^{}_{l} Y^{\dagger}_{l} \right) \;, \label{3}
\end{eqnarray}
where $g^{}_{i}$ with $i=1,2,3$ are the gauge couplings and satisfy the following RGEs
\begin{eqnarray}
16 \pi^2 \frac{{\rm d} g^{}_{i} }{{\rm d}t} = f^{}_{i} g^3_i \;,\label{4}
\end{eqnarray}
where $\left( f^{}_{1},f^{}_{2},f^{}_{3} \right)=\left( 41/10,-19/6,-7 \right)$ in the SM or $\left( 33/5,1,-3 \right)$ in the MSSM.
By defining four Hermitian matrices $H^{}_{\rm f} = M^{}_{\rm f} M^{\dagger}_{\rm f}$, Eq. (\ref{1}) can be rewritten as 
\begin{eqnarray}
&& 16 \pi^2 \frac{{\rm d} H^{}_{\rm u}}{{\rm d}t} = 2 \alpha^{}_{\rm u} H^{}_{\rm u} 
+ \frac{3b}{2} \left[ \left( Y^{}_{\rm u} Y^{\dagger}_{\rm u} \right) H^{}_{\rm u} + H^{}_{\rm u} \left( Y^{}_{\rm u} Y^{\dagger}_{\rm u} \right) \right] \;,\nonumber \\
&& 16 \pi^2 \frac{{\rm d} H^{}_{\rm d}}{{\rm d}t} = 2 \alpha^{}_{\rm d} H^{}_{\rm d} 
+ \frac{3c}{2} \left[ \left( Y^{}_{\rm u} Y^{\dagger}_{\rm u} \right) H^{}_{\rm d} + H^{}_{\rm d} \left( Y^{}_{\rm u} Y^{\dagger}_{\rm u} \right) \right] \;,\nonumber \\
&& 16 \pi^2 \frac{{\rm d} H^{}_{\nu}}{{\rm d}t} = 2 \alpha^{}_{\nu} H^{}_{\nu} 
+ \frac{3c}{2} \left[ \left( Y^{}_{l} Y^{\dagger}_{l} \right) H^{}_{\nu} + H^{}_{\nu} \left( Y^{}_{l} Y^{\dagger}_{l} \right) \right] \;,\nonumber \\
&& 16 \pi^2 \frac{{\rm d} H^{}_{l}}{{\rm d}t} = 2 \alpha^{}_{l} H^{}_{l} 
+ \frac{3b}{2} \left[ \left( Y^{}_l Y^{\dagger}_l \right) H^{}_l + H^{}_l \left( Y^{}_l Y^{\dagger}_l \right) \right] \;, \label{5}
\end{eqnarray}
where we have neglected the non-leading Yukawa matrices $Y^{}_{\rm d}$ and $Y^{}_{\nu}$.

In order to solve the above equations, we first diagonalize $H^{}_{\rm u}$ and $H^{}_{l}$ with two unitary matrices $O^{}_{\rm u}$ and $O^{}_{l}$ as follows
\begin{eqnarray}
O^\dagger_{\rm u} H^{}_{\rm u} O^{}_{\rm u} = D^{}_{\rm u} \equiv {\rm Diag} \left( m^{2}_{\rm u}, m^{2}_{\rm c}, m^{2}_{t}\right) \;,\quad 
O^\dagger_{l} H^{}_{l} O^{}_{l} = D^{}_{l} \equiv {\rm Diag} \left( m^{2}_{e}, m^{2}_{\mu}, m^{2}_{\tau}\right) \;. 
\end{eqnarray}
Then we redefine two effective mass matrices
\begin{eqnarray}
D^{}_{\rm d} \equiv O^\dagger_{\rm u} H^{}_{\rm d} O^{}_{\rm u} \;,\quad D^{}_{\nu} \equiv O^\dagger_{l} H^{}_{\nu} O^{}_{l} \;. \label{6}
\end{eqnarray}
$D^{}_{\rm d}$ and $D^{}_{\nu}$ are in general non-diagonal.
Since a Hermitian matrix $H^{}_{\rm u}$(or $H^{}_{l}$) totally has nine parameters, given three eigenvalues in $D^{}_{\rm u}$(or $D^{}_{l}$), 
$O^{}_{\rm u}$(or $O^{}_{l}$) consists of six free parameters and can be parametrized as $O^{}_{i} = P^{}_{i}  U^{}_{i} $ with $i = {\rm u},l$ and 
\begin{eqnarray}
P^{}_{i} \equiv \begin{pmatrix} 1 && \\ & e^{ {\rm i} \phi^{i}_{2} } & \\ & & e^{ {\rm i} \phi^{i}_{3} } \end{pmatrix} \;, \quad 
U^{}_{i} \equiv \begin{pmatrix} 1 & 0 & 0 \\ 0 & c^{i}_{23} & s^{i}_{23} \\ 0 & - s^{i}_{23} & c^{i}_{23} \end{pmatrix}
\begin{pmatrix} c^{i}_{13} & 0 & s^{i}_{13} e^{- {\rm i} \delta^{i} } \\ 0 & 1 & 0 \\ -  s^{i}_{13} e^{ {\rm i} \delta^{i} } & 0 & c^{i}_{13} \end{pmatrix}
\begin{pmatrix} c^{i}_{12} & s^{i}_{12} & 0 \\ - s^{i}_{12} & c^{i}_{12} & 0 \\ 0 & 0 & 1 \end{pmatrix} \;. \label{7}
\end{eqnarray}
We consider the running behavior of $H^{}_{\rm u}$, firstly. By differentiating $D^{}_{\rm u}$, one obtain 
\begin{eqnarray}
&& \frac{{\rm d} D^{}_{\rm u}}{{\rm d}t} = \dot O^\dagger_{\rm u} O^{}_{\rm u} D^{}_{\rm u} + O^\dagger_{\rm u} \dot H^{}_{\rm u} O^{}_{\rm u} + D^{}_{\rm u} O^\dagger_{\rm u} \dot O^{}_{\rm u} \;.\label{8} 
\end{eqnarray}
With the help of the definition $T^{}_{\rm u} \equiv O^\dagger_{\rm u} \dot O^{}_{\rm u} $ and Eq. (\ref{5}), Eq. (\ref{8}) leads us to
\begin{eqnarray}
\frac{{\rm d} D^{}_{\rm u}}{{\rm d}t} = - T^{}_{\rm u} D^{}_{\rm u} + D^{}_{\rm u} T^{}_{\rm u} 
+ \frac{1}{16 \pi^2} \left[ 2 \alpha^{}_{\rm u} D^{}_{\rm u} + \frac{3b}{2} y^{2}_{t} \left( E^{}_{3} D^{}_{\rm u} + D^{}_{\rm u} E^{}_{3} \right) \right] \;, \label{9}
\end{eqnarray}
where
\begin{eqnarray}
E^{}_{3} ={\rm Diag} \left( 0 , 0 , 1 \right) \;. \label{10}
\end{eqnarray}
To get instructive analytical results, we constrain the ratio of Higgs vacuum expectation values to be small enough $\tan \beta \leqslant 10$~\cite{Xing:1996hi},
so that the contributions of $y^{}_{u}$ and $y^{}_{c}$ in coefficients can be safely neglected since $y^{}_{u} \ll y^{}_{ c} \ll y^{}_{t}$.
The diagonal part of Eq. (\ref{9}) gives the RGEs of the eigenvalues
\begin{eqnarray}
16 \pi^2 \frac{{\rm d} D^{}_{\rm u} }{{\rm d}t} = 2 \alpha^{}_{\rm u} D^{}_{\rm u} 
+ \frac{3b}{2} y^{2}_{t} \left( E^{}_{3} D^{}_{\rm u} + D^{}_{\rm u} E^{}_{3} \right) \;. \label{11} 
\end{eqnarray}
The off-diagonal parts of Eq. (\ref{9}) can be extracted as
\begin{eqnarray}
- T^{ij}_{\rm u} D^{j}_{\rm u} + D^{i}_{\rm u} T^{ij}_{\rm u}  = 0 \quad\Rightarrow\quad T^{ij}_{\rm u}  = 0 \;, \label{12}
\end{eqnarray}
here $i\neq j$ and $D^{i}_{\rm u} \neq D^{j}_{\rm u}$, since the mass spectra of quark are non-degenerate. 
 After substituting  $O^{}_{\rm u} = P^{}_{\rm u} U^{}_{\rm u} $ into $T^{}_{\rm u}$, we obtain
\begin{eqnarray}
T^{}_{\rm u} = O^\dagger_{\rm u} \dot O^{}_{\rm u} 
= U^\dagger_{\rm u} P^\dagger_{\rm u} \left( \dot P^{}_{\rm u} U^{}_{\rm u} + P^{}_{\rm u} \dot U^{}_{\rm u} \right) 
= U^\dagger_{\rm u} \dot U^{}_{\rm u} + U^\dagger_{\rm u} P^\dagger_{\rm u} \dot P^{}_{\rm u} U^{}_{\rm u} \;. \label{13}
\end{eqnarray}                                                                                                   
Then the explicit forms of $T^{ij}_{\rm u} = 0 ~(i\neq j)$ is 
\begin{eqnarray}
&& \hspace{-1cm} \begin{pmatrix}
0 & 1 & s^{}_{13} c^{}_{\delta } & 0 & -c^{}_{23} s^{}_{13} s^{}_{23} s^{}_{\delta } & c^{}_{23} s^{}_{13} s^{}_{23} s^{}_{\delta } \\
0 & 0 & - s^{}_{13} s^{}_{\delta } \cos 2 \theta^{}_{12} & c^{}_{12} s^{}_{12} s_{13}^2 & A^{}_{25} & A^{}_{26} \\
c^{}_{12} c^{}_{\delta } & 0 & -s^{}_{12} c^{}_{13} & -c^{}_{12} c^{}_{13} s^{}_{13} s^{}_{\delta } & -c^{}_{12} c^{}_{13} s^{}_{13} s_{23}^2 s^{}_{\delta} 
& -c^{}_{12} c^{}_{13} s^{}_{13} c_{23}^2 s^{}_{\delta } \\
-c^{}_{12} s^{}_{\delta } & 0 & 0 & -c^{}_{12} c^{}_{13} s^{}_{13} c^{}_{\delta } & A^{}_{45} & A^{}_{46} \\
s^{}_{12} c^{}_{\delta } & 0 & c^{}_{12} c^{}_{13} & - s^{}_{12} c^{}_{13} s^{}_{13} s^{}_{\delta } & - s^{}_{12} c^{}_{13} s^{}_{13} s_{23}^2 s^{}_{\delta} 
& - s^{}_{12} c^{}_{13} s^{}_{13} c_{23}^2 s^{}_{\delta } \\
-s^{}_{12} s^{}_{\delta } & 0 & 0 & - s^{}_{12} c^{}_{13} s^{}_{13} c^{}_{\delta } & A^{}_{65} & A^{}_{66}
\end{pmatrix} 
\begin{pmatrix}
\dot \theta^{}_{13} \\ \dot \theta^{ }_{12} \\ \dot \theta^{ }_{23} \\ \dot \delta^{ } \\ \dot \phi^{ }_{2}\\ \dot \phi^{ }_{3}
\end{pmatrix} = 0  \;, \label{14}
\end{eqnarray}
with
\begin{eqnarray}
A^{}_{25} &=& c^{}_{12} s^{}_{12}\left(s_{13}^2 s_{23}^2 - c_{23}^2\right) - \frac{1}{2} s^{}_{13} c^{}_{\delta } \cos 2 \theta^{}_{12} \sin 2 \theta^{}_{23}  \;,\nonumber \\
A^{}_{26} &=& c^{}_{12} s^{}_{12}\left(s_{13}^2 c_{23}^2 - s_{23}^2\right)+ \frac{1}{2} s^{}_{13} s^{}_{\delta } \cos 2 \theta^{}_{12} \sin 2 \theta^{}_{23} \;,\nonumber \\
A^{}_{45} &=& -\frac{1}{2} c^{}_{13} \left(s^{}_{12} \sin 2 \theta^{}_{23} + 2 c^{}_{12} s^{}_{13} s_{23}^2 c^{}_{\delta } \right)  \;,\nonumber \\
A^{}_{46} &=&\hspace{0.3cm} \frac{1}{2} c^{}_{13} \left(s^{}_{12} \sin 2 \theta^{}_{23} - 2 c^{}_{12} s^{}_{13} c_{23}^2 c^{}_{\delta } \right)  \;,\nonumber \\
A^{}_{65} &=&\hspace{0.3cm} \frac{1}{2} c^{}_{13} \left(c^{}_{12} \sin 2 \theta^{}_{23} - 2 s^{}_{12} s^{}_{13} s_{23}^2 c^{}_{\delta } \right)  \;,\nonumber \\
A^{}_{66} &=& -\frac{1}{2} c^{}_{13} \left(c^{}_{12} \sin 2 \theta^{}_{23} + 2 s^{}_{12} s^{}_{13} c_{23}^2 c^{}_{\delta } \right)  \;.  \nonumber 
\end{eqnarray}
Solving Eq. (\ref{14}) directly, we can obtain 
\begin{eqnarray}
\dot \phi^{\rm u}_{2} = \dot \phi^{\rm u}_{3} = \dot \theta^{\rm u}_{12} = \dot \theta^{\rm u}_{13} = \dot \theta^{\rm u}_{23} = \dot \delta^{\rm u}= 0 \;. \label{15}
\end{eqnarray}
That means the unitary matrix $O^{}_{\rm u}$ does not evolve with energy scale. But the three masses get modified at different energy scale according to Eq. (\ref{11}).
If $H^{}_{\rm u}$ is chosen as diagonal at initial scale, then $H^{}_{\rm u}$ keeps diagonal during the RGE running. This scenario is employed in most literatures.

After integrating Eq. (\ref{11}) from $t^{}_{0} \equiv \ln (\Lambda/\Lambda^{}_{\rm EW})$ with a high energy scale $\Lambda$ where the mass matrices of fermion may have some simple structures,
down to $t=0$ (i.e., the electroweak scale $\Lambda^{}_{\rm EW}$), we can obtain the eigenvalues of up-type quark matrix at $\Lambda^{}_{\rm EW}$, 
which are denoted as $D^{}_{\rm u} \left(t\right)$ 
\begin{eqnarray}                                                                                                            
D^{}_{\rm u} \left(t\right) = \gamma^{2}_{\rm u} I^{}_{\rm u} D^{}_{\rm u} \left(t^{}_{0}\right) I^{}_{\rm u} \;, \label{16}
\end{eqnarray}
where 
\begin{eqnarray}
\gamma^{}_{\rm u} = \exp \left( \frac{1}{16\pi^2} \int^{t}_{t^{}_{0}} \alpha^{}_{\rm u} {\rm d}t' \right) \;,\quad 
\xi^{}_{t} = \exp \left( \frac{1}{16\pi^2} \int^{t}_{t^{}_{0}} y^{2}_{t} {\rm d}t' \right) \;,\quad 
I^{}_{\rm u} = {\rm Diag} \left( 1,1, \xi^{3b/2}_{t} \right) \;. \label{17}
\end{eqnarray}
The magnitudes of $\gamma^{}_{\rm u}$ and $\xi^{}_{t}$ for different energy scale $\Lambda$ are shown in figure 1, where the running quark and charged-lepton masses and other SM parameters renormalized to the energy scale $\mu=M^{}_{Z}$ have been input \cite{Xing:2011aa,Tanabashi:2018oca}.

Transforming $D^{}_{\rm u} \left(t\right)$ into $H^{}_{\rm u} \left(t\right)$, we finally obtain 
\begin{eqnarray}
H^{}_{\rm u} \left(t\right) = O^{}_{\rm u} D^{}_{\rm u} \left(t\right) O^{\dagger}_{\rm u}
= \gamma^{2}_{\rm u} O^{}_{\rm u} I^{}_{\rm u} O^{\dagger}_{\rm u} H^{}_{\rm u} \left(t^{}_{0}\right) O^{}_{\rm u} I^{}_{\rm u} O^{\dagger}_{\rm u} \;, \label{18}
\end{eqnarray}
here we have used the previous result that $O^{}_{\rm u}$ is independent of scale.
Obviously, $H^{}_{\rm u} \left(t\right)$ is still a Hermitian matrix. The RGE-corrected mass matrix $M^{}_{\rm u} \left(t\right)$ can be 
extracted from Eq. (\ref{18})
\begin{eqnarray}
M^{}_{\rm u} \left(t\right) = \gamma^{}_{\rm u} O^{}_{\rm u} I^{}_{\rm u} O^{\dagger}_{\rm u} M^{}_{\rm u} \left(t^{}_{0}\right) \;. \label{19}
\end{eqnarray}
Actually, the general solution is $M^{}_{\rm u} \left(t\right) V$ with an arbitrary unitary matrix $V$ which can be absorbed by rotating the right-handed quark fields.
Then we only consider the case of Eq. (\ref{19}). 
If we choose a special basis in which $M^{}_{\rm u}\left(t^{}_{0}\right)$ is Hermitian at the energy scale $t^{}_{0}$, 
i.e. $ O^{\dagger}_{\rm u} M^{}_{\rm u}\left(t^{}_{0}\right) O^{}_{\rm u} = \hat M^{}_{\rm u}\left(t^{}_{0}\right) 
\equiv {\rm Diag} \left( m^{}_{\rm u}, m^{}_{\rm c}, m^{}_{t}\right)$, then $M^{}_{\rm u} \left(t\right)$ is also Hermitian
\begin{eqnarray}
M^{\dagger}_{\rm u} \left(t\right) = \gamma^{}_{\rm u} M^{}_{\rm u} \left(t^{}_{0}\right) O^{}_{\rm u} I^{}_{\rm u} O^{\dagger}_{\rm u}
= \gamma^{}_{\rm u} O^{}_{\rm u} \hat M^{}_{\rm u} \left(t^{}_{0}\right) I^{}_{\rm u} O^{\dagger}_{\rm u}
= \gamma^{}_{\rm u} O^{}_{\rm u} I^{}_{\rm u} O^{\dagger}_{\rm u} M^{}_{\rm u} \left(t^{}_{0}\right)
= M^{}_{\rm u} \left(t\right) \;. \label{20}
\end{eqnarray}

Then we proceed to consider the running behavior of $H^{}_{\rm d}$. By differentiating $D^{}_{\rm d}$, one obtain 
\begin{eqnarray}
\frac{{\rm d} D^{}_{\rm d}}{{\rm d}t} = O^\dagger_{\rm u} \frac{{\rm d} H^{}_{\rm d}}{{\rm d}t} O^{}_{\rm u}
= \frac{1}{16 \pi^2} \left[ 2 \alpha^{}_{\rm d} D^{}_{\rm d} + \frac{3c}{2} y^2_t \left( E^{}_{3} D^{}_{\rm d} + D^{}_{\rm d} E^{}_{3} \right) \right] \;,\label{21}
\end{eqnarray}
where we have used Eq. (\ref{5}) and $\dot O^{}_{\rm u} = 0$. After integrating Eq. (\ref{21}), we can get the evolved $D^{}_{\rm d}$ 
\begin{eqnarray}
D^{}_{\rm d}\left(t\right) = \gamma^{2}_{\rm d} I^{}_{\rm d} D^{}_{\rm d} \left(t^{}_{0}\right) I^{}_{\rm d} \;, \label{22}
\end{eqnarray}
where $\xi^{}_{t}$ is given by Eq. (\ref{17}) and 
\begin{eqnarray}
\gamma^{}_{\rm d} = \exp \left( \frac{1}{16\pi^2} \int^{t}_{t^{}_{0}} \alpha^{}_{\rm d} {\rm d}t \right) \;,\quad 
I^{}_{\rm d} = {\rm Diag} \left( 1,1, \xi^{3c/2}_{t}\right)\;. \label{23}
\end{eqnarray}
The magnitudes of $\gamma^{}_{\rm d}$ for different energy scale $\Lambda$ are shown in figure 1.
The RGE-corrected matrices $H^{}_{\rm d} \left(t\right)$ and $M^{}_{\rm d} \left(t\right)$ are given by
\begin{eqnarray}
H^{}_{\rm d}\left(t\right) &=& \gamma^{2}_{\rm d} O^{}_{\rm u} I^{}_{\rm d} O^\dagger_{\rm u} H^{}_{\rm d}\left(t^{}_{0}\right) O^{}_{\rm u} I^{}_{\rm d} O^\dagger_{\rm u} \;, \label{24}\\
M^{}_{\rm d}\left(t\right) &=& \gamma^{}_{\rm d} O^{}_{\rm u} I^{}_{\rm d} O^\dagger_{\rm u} M^{}_{\rm d}\left(t^{}_{0}\right) \;. \label{25}
\end{eqnarray}
After a brief check, we can see that $H^{}_{\rm d}\left(t\right)$ is still Hermitian, but $M^{}_{\rm d}\left(t\right)$ is not Hermitian any more.
\begin{figure}[t]
\begin{center}
\includegraphics[scale=0.5]{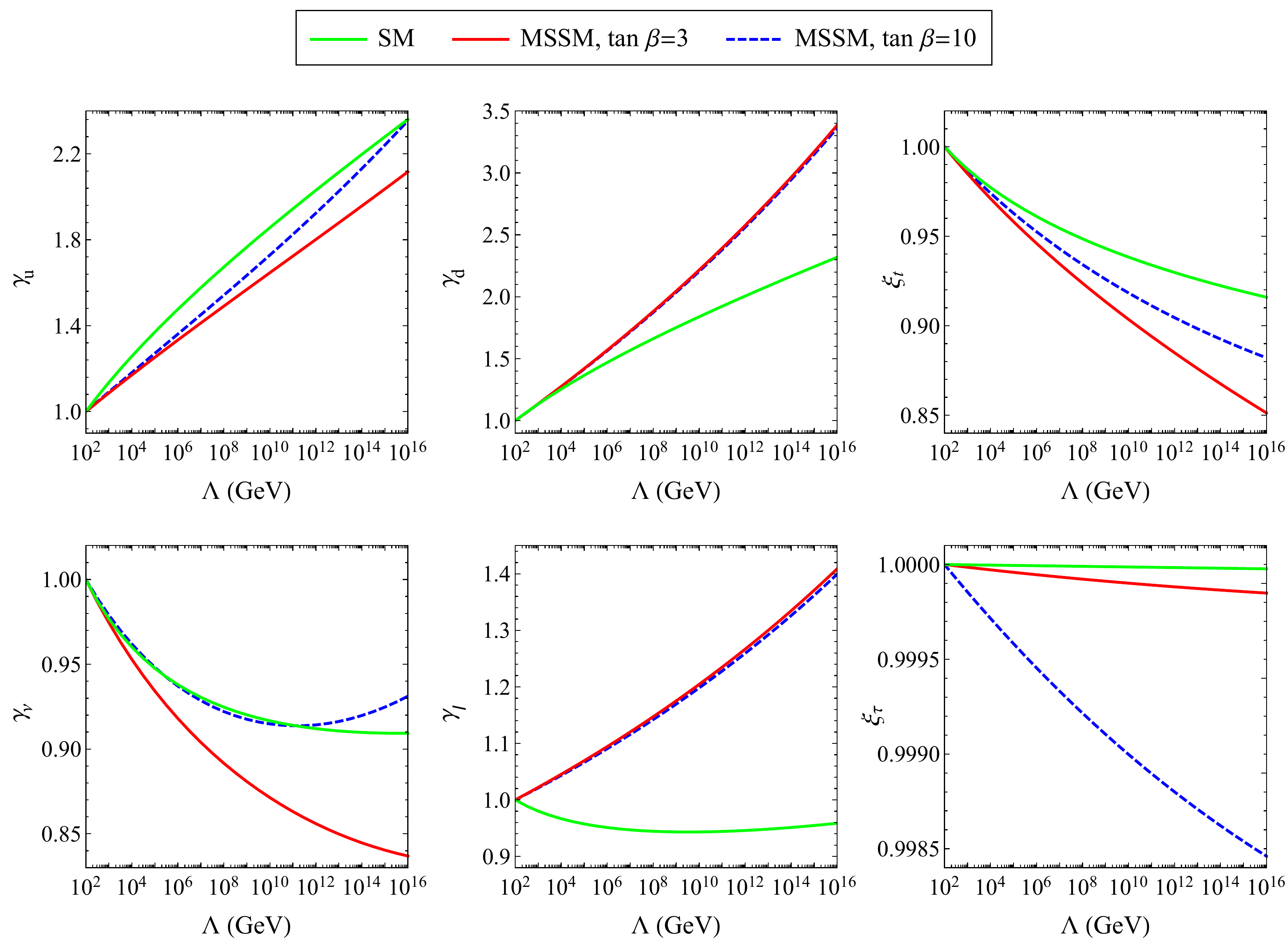} 	
\caption{In the case of Dirac neutrino, illustration for the renormalization-group evolutions of $\gamma^{}_{\rm u}$, $\gamma^{}_{\rm d}$ and $\xi^{}_{t}$ in the quark sector (the top three figures) 
and $\gamma^{}_{\nu}$, $\gamma^{}_{l}$ and $\xi^{}_{\tau}$ in the lepton sector  (the bottom three figures) for different energy scale $\Lambda$, where the running quark and charged-lepton masses and other SM parameters renormalized to the energy scale $\mu=M^{}_{Z}$ have been input \cite{Xing:2011aa,Tanabashi:2018oca}. } 
\end{center}
\end{figure}

In the lepton sector, the calculation is similar. Here we list the main result,
\begin{itemize}
	\item The unitary matrix $O^{}_{l}$ is independent of energy scale, $\dot O^{}_{l} = 0$.
	\item The evolved $H^{}_{l}\left(t\right)$ and $H^{}_{\nu}\left(t\right)$ are given by
	\begin{eqnarray}
	&&  H^{}_{l}\left(t\right) = \gamma^{2}_{l} O^{}_{l} I^{}_{l} O^{\dagger}_{l} H^{}_{l} \left(t^{}_{0}\right) O^{\dagger}_{l} I^{}_{l} O^{\dagger}_{l} \;, \label{26} \\
	&& H^{}_{\nu}\left(t\right) = \gamma^{2}_{\nu} O^{}_{l} I^{}_{\nu} O^{\dagger}_{l} H^{}_{\nu} \left(t^{}_{0}\right) O^{\dagger}_{l} I^{}_{\nu} O^{\dagger}_{l} \;, \label{27} 
	\end{eqnarray}
	with 
	\begin{eqnarray}
	&& \gamma^{}_{l,\nu} = \exp \left( \frac{1}{16\pi^2} \int^{t}_{t^{}_{0}} \alpha^{}_{l,\nu} {\rm d}t' \right) \;,\quad 
	\xi^{}_{\tau} = \exp \left( \frac{1}{16\pi^2} \int^{t}_{t^{}_{0}} y^{2}_{\tau} {\rm d}t' \right) \;,\nonumber \\
	&& I^{}_{l} = {\rm Diag} \left( 1,1, \xi^{3b/2}_{\tau}\right)\;,\quad 
	I^{}_{\nu} = {\rm Diag} \left( 1,1, \xi^{3c/2}_{\tau}\right)\;. \label{28}
	\end{eqnarray}
	Here the contributions of $y^{}_{e}$ and $y^{}_{\mu}$ are neglected because of $y^{}_{e} \ll y^{}_{\mu} \ll y^{}_{\tau}$. 
	The magnitudes of $\gamma^{}_{\nu}$, $\gamma^{}_{l}$ and $\xi^{}_{\tau}$ for different energy scale $\Lambda$ are shown in figure 1.
	\item The evolved mass matrices $M^{}_{l}\left(t\right)$ and $M^{}_{\nu}\left(t\right)$ are given by
	\begin{eqnarray}
	&&  M^{}_{l}\left(t\right) = \gamma^{}_{l} O^{}_{l} I^{}_{l} O^{\dagger}_{l} M^{}_{l} \left(t^{}_{0}\right) \;, \label{29} \\
	&& M^{}_{\nu}\left(t\right) = \gamma^{}_{\nu} O^{}_{l} I^{}_{\nu} O^{\dagger}_{l} M^{}_{\nu} \left(t^{}_{0}\right) \;. \label{30} 
	\end{eqnarray}
	\item If $M^{}_{l} \left(t^{}_{0}\right)$ and $M^{}_{\nu} \left(t^{}_{0}\right)$ are chose as Hermitian, then $M^{}_{l}\left(t\right)$ is still Hermitian, but $M^{}_{\nu}\left(t\right)$ is not.
\end{itemize}
The elements of mass matrices $M^{}_{\rm u}\left(t\right)$, $M^{}_{\rm d}\left(t\right)$, $M^{}_{l}\left(t\right)$ and $M^{}_{\nu}\left(t\right)$ are 
\begin{eqnarray}
M^{ij}_{\rm u} \left(t\right) &=& \gamma^{}_{\rm u} \left[ M^{ij}_{\rm u} \left(t^{}_{0}\right) 
+ \left( \xi^{3b/2}_{t} - 1 \right) \sum^{3}_{k=1} O^{i3}_{\rm u} O^{k3*}_{\rm u} M^{kj}_{\rm u} \left(t^{}_{0}\right) 
\right] \;, \nonumber \\
M^{ij}_{\rm d} \left(t\right) &=& \gamma^{}_{\rm d} \left[ M^{ij}_{\rm d} \left(t^{}_{0}\right) 
+ \left( \xi^{3c/2}_{t} - 1 \right) \sum^{3}_{k=1} O^{i3}_{\rm u} O^{k3*}_{\rm u} M^{kj}_{\rm d} \left(t^{}_{0}\right) 
\right] \;, \nonumber \\
M^{ij}_{\nu} \left(t\right) &=& \gamma^{}_{\nu} \left[ M^{ij}_{\nu} \left(t^{}_{0}\right) 
+ \left( \xi^{3c/2}_{\tau} - 1 \right) \sum^{3}_{k=1} O^{i3}_{l} O^{k3*}_{l} M^{kj}_{\nu} \left(t^{}_{0}\right) 
\right] \;, \nonumber \\
M^{ij}_{l} \left(t\right) &=& \gamma^{}_{l} \left[ M^{ij}_{l} \left(t^{}_{0}\right) 
+ \left( \xi^{3b/2}_{\tau} - 1 \right) \sum^{3}_{k=1} O^{i3}_{l} O^{k3*}_{l} M^{kj}_{l} \left(t^{}_{0}\right) 
\right] \;.\label{31}
\end{eqnarray}
Eqs. (\ref{19}), (\ref{25}), (\ref{29}) and (\ref{30}) show that, the overall factors $\gamma^{}_{\rm f}$ (for ${\rm f = u,d,}l,\nu$) only affect the magnitudes of the fermion masses, 
while $I^{}_{\rm f}$ and $O^{}_{{\rm u},l}$ can modify the structures of fermion mass matrices. 
Furthermore, only $O^{}_{{\rm u},l}$, or more specifically the third column of $O^{}_{{\rm u},l}$ , may effect the texture zeros of fermion mass matrices, since $I^{}_{\rm f}$ is a diagonal matrix. If we choose the basis of $M^{}_{\rm u}$ and $M^{}_{l}$ being diagonal (i.e. $O^{}_{\rm u}=O^{}_{l}=1$) at initial scale $t^{}_{0}$, 
$M^{}_{\rm u}$ and $M^{}_{l}$ remain diagonal at any scale, and the possible texture zeros of $M^{}_{\rm d}$ and $M^{}_{\nu}$ are essentially stable against RGE running effect.
In general case, when $M^{}_{{\rm u},l}\left(t^{}_{0}\right) $ (i.e. $O^{}_{{\rm u},l}$) have some simple structures, the possible texture zeros of some fermion mass matrices
could be invariant with scale in a reliable approximation.
\subsection{The case of Majorana neutrinos}
The fermion mass terms in SM and MSSM are given by
\begin{eqnarray}
\mathcal{L}^{}_{\rm SM} &\supset& \frac{v}{\sqrt 2} \bar u^{}_{L} Y^{}_{\rm u} u^{}_{R} + \frac{v}{\sqrt 2} \bar d^{}_{L} Y^{}_{\rm u} d^{}_{R}
+ \frac{1}{2} \bar \nu^{}_{L} M^{}_{\nu} \nu^{c}_{L} + \frac{v}{\sqrt 2} \bar l^{}_{L} Y^{}_{l} l^{}_{R} +  {\rm h.c.}\;, \nonumber \\
\mathcal{L}^{}_{\rm MSSM} &\supset& \frac{v \sin \beta }{\sqrt 2} \bar u^{}_{L} Y^{}_{\rm u} u^{}_{R} + \frac{v \cos \beta }{\sqrt 2} \bar d^{}_{L} Y^{}_{\rm u} d^{}_{R}
+ \frac{1}{2} \bar \nu^{}_{L} M^{}_{\nu} \nu^{c}_{L} + \frac{v \cos \beta }{\sqrt 2} \bar l^{}_{L} Y^{}_{l} l^{}_{R} +  {\rm h.c.} \;.
\end{eqnarray}
The one-loop renormalization-group equations for fermion mass matrices are given by \cite{Xing:2011zza}
\begin{eqnarray}
16 \pi^2 \frac{{\rm d}M^{}_{\nu}}{{\rm d}t} &=& \alpha^{}_{\nu} M^{}_{\nu} + C^{}_{\nu} \left[ \left(Y^{}_{l} Y^{\dagger}_{l}\right) M^{}_{\nu} + M^{}_{\nu} \left(Y^{}_{l} Y^{\dagger}_{l}\right)^T \right]
\;, \nonumber \\
16 \pi^2 \frac{{\rm d}M^{}_{l}}{{\rm d}t} &=& \Big[ \alpha^{}_{l} + \frac{3 b}{2} Y^{}_{l} Y^{\dagger}_{l} \Big] M^{}_{l}
\;, \nonumber \\
16 \pi^2 \frac{{\rm d}M^{}_{\rm u}}{{\rm d}t} &=& \Big[ \alpha^{}_{\rm u} + \frac{3}{2} \left( b Y^{}_{\rm u} Y^{\dagger}_{\rm u} + c Y^{}_{\rm d} Y^{\dagger}_{\rm d} \right) \Big] 
M^{}_{\rm u} \;, \nonumber \\
16 \pi^2 \frac{{\rm d}M^{}_{\rm d}}{{\rm d}t} &=& \Big[ \alpha^{}_{\rm d} + \frac{3}{2} \left( c Y^{}_{\rm u} Y^{\dagger}_{\rm u} + b Y^{}_{\rm d} Y^{\dagger}_{\rm d} \right) \Big] 
M^{}_{\rm d} \;, \label{32}
\end{eqnarray}
where $\alpha^{}_{\rm f} = G^{}_{\rm f} + R^{}_{\rm f}$ with ${\rm f=u,d},\nu,l$. The quantities $R^{}_{\rm f}$ are given as follows
\begin{eqnarray}
&& R^{}_{\rm u} = R^{}_{\nu} /2 = {\rm Tr} \left( 3 Y^{}_{\rm u} Y^{\dagger}_{\rm u} + 3a Y^{}_{\rm d} Y^{\dagger}_{\rm d} + a Y^{}_{l} Y^{\dagger}_{l} \right) \;, \nonumber \\
&& R^{}_{\rm d} = R^{}_{l} = {\rm Tr} \left( 3a Y^{}_{\rm u} Y^{\dagger}_{\rm u} + 3 Y^{}_{\rm d} Y^{\dagger}_{\rm d} + Y^{}_{l} Y^{\dagger}_{l} \right) \;. \label{33}
\end{eqnarray}
In the framework of the SM, we have 
\begin{eqnarray}
&& a = b = -c = 1 \;,\hspace{2cm} C^{}_{\nu} =  -3/2  \;,\nonumber \\
&& G^{}_{\nu} = - 3 g^2_2 + \lambda \;, \hspace{2cm} G^{}_{l} = - \frac{9}{4} g^2_1 - \frac{9}{4} g^2_2 \;, \nonumber \\
&& G^{}_{\rm u} = - \frac{17}{20} g^2_1 - \frac{9}{4} g^2_2 - 8 g^2_2 \;,\quad G^{}_{\rm d} = - \frac{1}{4} g^2_1 - \frac{9}{4} g^2_2 - 8 g^2_2 \;, \label{34}
\end{eqnarray}
and in the framework of the MSSM we have 
\begin{eqnarray}
&& a = 0 \;,\qquad b = 2 \;,\qquad c = 2/3 \;,\qquad C^{}_{\nu} = 1 \;,\nonumber \\
&& G^{}_{\nu} = - \frac{6}{5} g^2_1 - 6 g^2_2 \;, \hspace{2cm} G^{}_{l} = - \frac{9}{5} g^2_1 - 3 g^2_2 \;, \nonumber \\
&& G^{}_{\rm u} = - \frac{13}{15} g^2_1 - 3 g^2_2 - \frac{16}{3} g^2_3 \;,\quad G^{}_{\rm d} = - \frac{7}{15} g^2_1 - 3 g^2_2 - \frac{16}{3} g^2_2 \;. \label{35}
\end{eqnarray}
The RGEs of the three gauge couplings $g^{}_{1}$, $g^{}_{2}$ and $g^{}_{3}$ are Eq. (\ref{4}),  $\lambda$ is the Higgs self-coupling paramter of the SM
and obeys the RGE
\begin{eqnarray}
&& 16 \pi^{2} \frac{{\rm d} \lambda}{\rm{d} t}=6 \lambda^{2} - 3 \lambda \left( \frac{3}{5} g_{1}^{2} + 3 g_{2}^{2} \right)
+ \frac{3}{2} \left( \frac{3}{5} g_{1}^{2} + g_{2}^{2} \right)^{2} + 3 g_{2}^{4}
\nonumber \\
&& \hspace{1.6cm} + 4 \lambda {\rm Tr} \left[ 3 \left(Y_{\rm u} Y_{\rm u }^{\dagger} \right) 
+ 3 \left( Y_{\rm d}^{} Y_{\rm d}^{\dagger} \right) + \left( Y^{}_{l} Y_{l}^{\dagger} \right) \right] \nonumber \\
&& \hspace{1.6cm} - 8 {\rm Tr} \left[ 3 \left(Y_{\rm u} Y_{\rm u }^{\dagger} \right)^2 
+ 3 \left( Y_{\rm d}^{} Y_{\rm d}^{\dagger} \right)^2 + \left( Y^{}_{l} Y_{l}^{\dagger} \right)^2 \right] \;. \label{36}
\end{eqnarray}
Neglecting all non-leading Yukawa couplings in the coefficients in Eq. (\ref{32}) and defining $H^{}_{{\rm u ,d}, l} = M^{}_{{\rm u ,d}, l} M^{\dagger}_{{\rm u ,d}, l}$,
Eq. (\ref{32}) can be rewritten as 
\begin{eqnarray}
16 \pi^2 \frac{{\rm d}M^{}_{\nu}}{{\rm d}t} &=& \alpha^{}_{\nu} M^{}_{\nu} 
+ C^{}_{\nu} \left[ \left(Y^{}_{l} Y^{\dagger}_{l}\right) M^{}_{\nu} + M^{}_{\nu} \left(Y^{}_{l} Y^{\dagger}_{l}\right)^T \right]
\;, \nonumber \\
16 \pi^2 \frac{{\rm d}H^{}_{l}}{{\rm d}t} &=& 2 \alpha^{}_{l} H^{}_{l} 
+ \frac{3b}{2} \left[ \left( Y^{}_{l} Y^{\dagger}_{l} \right) H^{}_{l} + H^{}_{l} \left( Y^{}_{l} Y^{\dagger}_{l} \right) \right]
\;, \nonumber \\
16 \pi^2 \frac{{\rm d}H^{}_{\rm u}}{{\rm d}t} &=& 2 \alpha^{}_{\rm u} H^{}_{\rm u} 
+ \frac{3b}{2} \left[ \left( Y^{}_{\rm u} Y^{\dagger}_{\rm u} \right) H^{}_{\rm u} + H^{}_{\rm u} \left( Y^{}_{\rm u} Y^{\dagger}_{\rm u} \right) \right]
\;, \nonumber \\
16 \pi^2 \frac{{\rm d}H^{}_{\rm d}}{{\rm d}t} &=& 2 \alpha^{}_{\rm d} H^{}_{\rm d} 
+ \frac{3c}{2} \left[ \left( Y^{}_{\rm u} Y^{\dagger}_{\rm u} \right) H^{}_{\rm d} + H^{}_{\rm d} \left( Y^{}_{\rm u} Y^{\dagger}_{\rm u} \right) \right]\;. \label{37}
\end{eqnarray}
Similar to the previous calculation, we introduce two unitary matrix $O^{}_{\rm u}$ and $O^{}_{l}$, which are independent of energy scale, to diagonalize $H^{}_{\rm u}$ and $H^{}_{l}$, respectively.
Then the evolved $H^{}_{l}$, $H^{}_{\rm u}$ and $H^{}_{\rm d}$ are given by
\begin{eqnarray}
H^{}_{\rm u} \left(t\right) &=& \gamma^{2}_{\rm u} O^{}_{\rm u} I^{}_{\rm u} O^{\dagger}_{\rm u} H^{}_{\rm u} \left(t^{}_{0}\right) O^{\dagger}_{\rm u} I^{}_{\rm u} O^{\dagger}_{\rm u}
\;, \nonumber \\
H^{}_{\rm d}\left(t\right) &=& \gamma^{2}_{\rm d} O^{}_{\rm u} I^{}_{\rm d} O^\dagger_{\rm u} H^{}_{\rm d}\left(t^{}_{0}\right) O^{}_{\rm u} I^{}_{\rm d} O^\dagger_{\rm u} \;,\nonumber \\
H^{}_{l}\left(t\right) &=& \gamma^{2}_{l} O^{}_{l} I^{}_{l} O^{\dagger}_{l} H^{}_{l} \left(t^{}_{0}\right) O^{\dagger}_{l} I^{}_{l} O^{\dagger}_{l} \;,\label{38}
\end{eqnarray}
the RGE-corrected mass matrices are 
\begin{eqnarray}
M^{}_{\rm u} \left(t\right) &=& \gamma^{}_{\rm u} O^{}_{\rm u} I^{}_{\rm u} O^{\dagger}_{\rm u} M^{}_{\rm u} \left(t^{}_{0}\right) \;, \nonumber \\
M^{}_{\rm d}\left(t\right) &=& \gamma^{}_{\rm d} O^{}_{\rm u} I^{}_{\rm d} O^\dagger_{\rm u} M^{}_{\rm d}\left(t^{}_{0}\right) \;,\nonumber \\
M^{}_{l}\left(t\right) &=& \gamma^{}_{l} O^{}_{l} I^{}_{l} O^{\dagger}_{l} M^{}_{l} \left(t^{}_{0}\right) \;, \nonumber \\
M^{}_{\nu}\left(t\right) &=& \gamma^{}_{\nu} O^{}_{l} I^{}_{\nu} O^{\dagger}_{l} M^{}_{\nu} \left(t^{}_{0}\right) O^{*}_{l} I^{}_{\nu} O^{T}_{l} \;,\label{39}
\end{eqnarray}
where 
\begin{eqnarray}
&& \gamma^{}_{{\rm u,d},\nu,l} = \exp \left( \frac{1}{16\pi^2} \int^{t}_{t^{}_{0}} \alpha^{}_{{\rm u,d},\nu,l} {\rm d}t \right) \;,\nonumber \\
&& \xi^{}_{\tau} = \exp \left( \frac{1}{16\pi^2} \int^{t}_{t^{}_{0}} y^{2}_{\tau} {\rm d}t \right) \;,\quad 
\xi^{}_{t} = \exp \left( \frac{1}{16\pi^2} \int^{t}_{t^{}_{0}} y^{2}_{t} {\rm d}t \right) \;,\nonumber \\
&& I^{}_{\rm u} = {\rm Diag} \left( 1,1, \xi^{3b/2}_{t}\right) \;,\hspace{1cm} I^{}_{\rm d} = {\rm Diag} \left( 1,1, \xi^{3c/2}_{t}\right) \;,\nonumber \\
&& I^{}_{\nu} = {\rm Diag} \left( 1,1, \xi^{C^{}_{\nu}}_{\tau} \right) \;,\hspace{1.4cm} I^{}_{l} = {\rm Diag} \left( 1,1, \xi^{3b/2}_{\tau}\right) \;.\label{40}
\end{eqnarray}
The magnitudes of ${\rm u,d},\nu,l$, $\xi^{}_{t}$ and $\xi^{}_{\tau}$ for different energy scale $\Lambda$ are shown in figure 2.
The elements of RGE-corrected mass matrices $M^{}_{\rm u}\left(t\right)$, $M^{}_{\rm d}\left(t\right)$, $M^{}_{l}\left(t\right)$ and $M^{}_{\nu}\left(t\right)$ are 
\begin{eqnarray}
M^{ij}_{\rm u} \left(t\right) &=& \gamma^{}_{\rm u} \left[ M^{ij}_{\rm u} \left(t^{}_{0}\right) 
+ \left( \xi^{3b/2}_{t} - 1 \right) \sum^{3}_{k=1} O^{i3}_{\rm u} O^{k3*}_{\rm u} M^{kj}_{\rm u} \left(t^{}_{0}\right) 
\right] \;, \nonumber \\
M^{ij}_{\rm d} \left(t\right) &=& \gamma^{}_{\rm d} \left[ M^{ij}_{\rm d} \left(t^{}_{0}\right) 
+ \left( \xi^{3c/2}_{t} - 1 \right) \sum^{3}_{k=1} O^{i3}_{\rm u} O^{k3*}_{\rm u} M^{kj}_{\rm d} \left(t^{}_{0}\right) 
\right] \;, \nonumber \\
M^{ij}_{l} \left(t\right) &=& \gamma^{}_{l} \left[ M^{ij}_{l} \left(t^{}_{0}\right) 
+ \left( \xi^{3b/2}_{\tau} - 1 \right) \sum^{3}_{k=1} O^{i3}_{l} O^{k3*}_{l} M^{kj}_{l} \left(t^{}_{0}\right) 
\right] \;, \nonumber \\
M^{ij}_{\nu} \left(t\right) &=& \gamma^{}_{\nu} \left\{
M^{ij}_{\nu} \left(t^{}_{0}\right) 
+ \left( \xi^{3c/2}_{\tau} - 1 \right) \sum^{3}_{k=1} O^{k3*}_{l} \left[ O^{i3}_{l} M^{kj}_{\nu} \left(t^{}_{0}\right) 
+ O^{j3}_{l} M^{ik}_{\nu} \left(t^{}_{0}\right) \right] \right\} \;,\label{41}
% \right. \nonumber \\
%&&\left. + \left( \xi^{3b/2}_{\tau} - 1 \right)^2 O^{i3}_{l} O^{j3}_{l}\sum^{3}_{k=1} M^{km}_{\nu} \left(t^{}_{0}\right) O^{k3*}_{l} O^{m3*}_{l} \right\} \;.
\end{eqnarray}
In the last equation, we have ignored the $\left( \xi^{3c/2}_{\tau} - 1 \right)^2$ term. As in the case of Dirac neutrinos, $I^{}_{\rm f}$ and $O^{}_{{\rm u},l}$ can modify the structure of fermion mass matrices, and only the third column of $O^{}_{{\rm u},l}$ may effect the texture zeros of fermion mass matrices.
\begin{figure}[t]
\begin{center}
\includegraphics[scale=0.5]{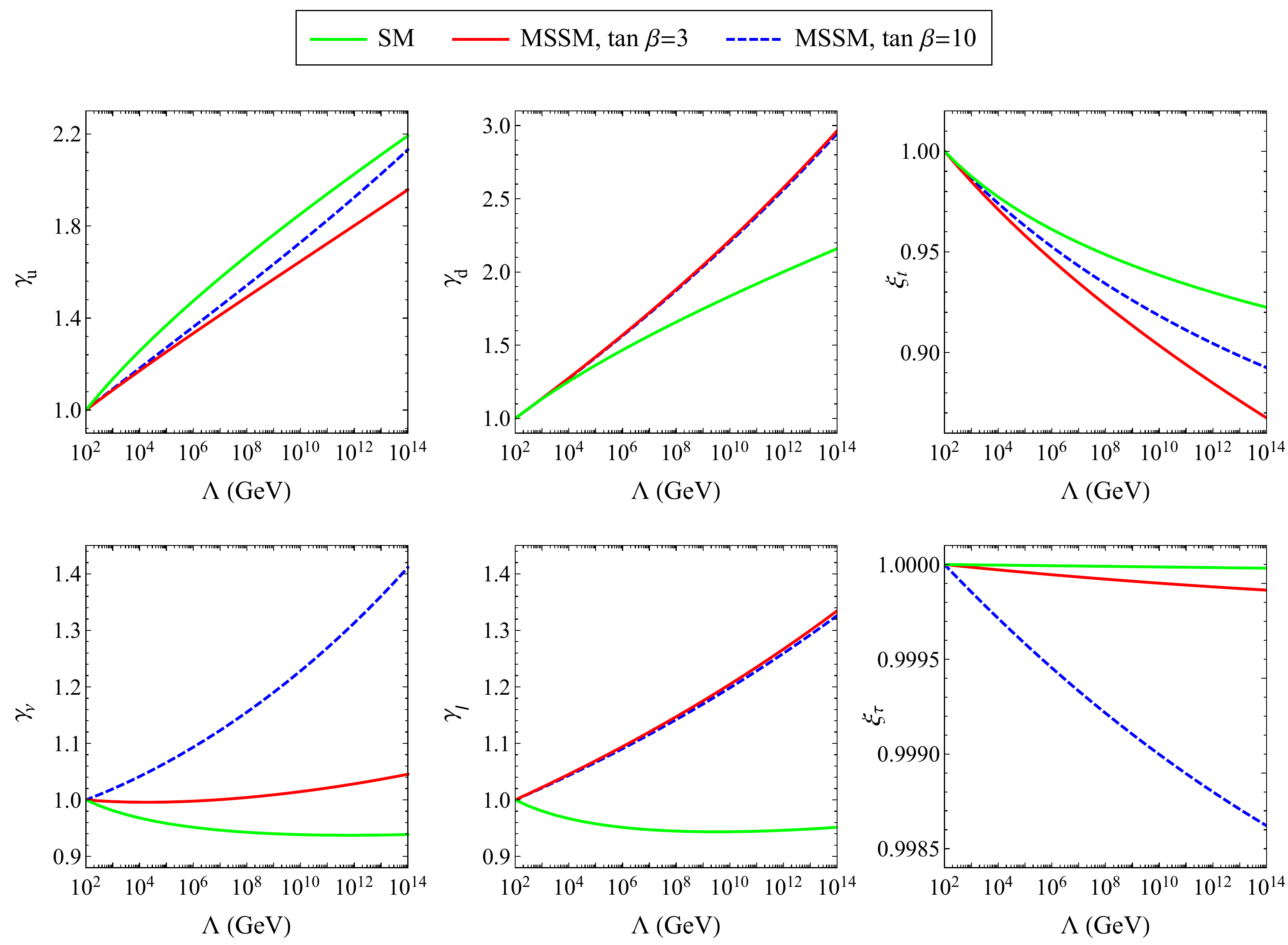} 	
\caption{In the case of Majorana neutrino, illustration for the renormalization-group evolutions $\gamma^{}_{\rm u}$, $\gamma^{}_{\rm d}$ and $\xi^{}_{t}$ in the quark sector (the top three figures) 
and $\gamma^{}_{\nu}$, $\gamma^{}_{l}$ and $\xi^{}_{\tau}$ in the lepton sector (the bottom three figures) for different energy scale $\Lambda$, where the running quark and charged-lepton masses and other SM parameters renormalized to the energy scale $\mu=M^{}_{Z}$ have been input \cite{Xing:2011aa,Tanabashi:2018oca}. } 
	\end{center}
\end{figure}
\section{Quark mass matrices with texture zeros}
As we know, the fermion mass matrices which come from the Higgs mechanism are arbitrary in the SM, therefore the number of free parameters is larger than 
the physical ones. For example, in the quark sector, there are 36 free parameters in up- and down-type quark mass matrices and only ten physical parameters corresponding to six quark masses,
three mixing angles and one CP phase. To get a predictable ansatz of fermion mass matrices, one has to reduce the free parameters.
Firstly, without loss of generality, the mass matrices $M^{}_{\rm u}$ and $M^{}_{\rm d}$ can be taken to be Hermitian in the SM or its extensions which have no 
flavor changing right-handed currents \cite{Fritzsch:1979zq}, since we can rotate freely the right-handed field which are singlets in such models.
Another way to limit the parameters is introducing a common weak-basis(WB) transformation to $M^{}_{\rm u}$ and $M^{}_{\rm d}$, 
this transformation has no physical content and can lead $M^{}_{\rm u}$ has two texture zeros and $M^{}_{\rm d}$ has one texture zero \cite{Fritzsch:1999rb, Branco:1999nb}. 
Any extra texture zero will achieve some phenomenological predictions of masses and mixing parameters.  
A typical example of this kind is the famous Fritzsch mass matrices \cite{Fritzsch:1977vd}.
In the lepton sector, Ref. \cite{Fukugita:1992sy} proposed a ansatz (hereafter FTY model) where neutrino mass are generated via type-I seesaw mechanism with the charged lepton 
and the Dirac neutrino mass matrices of Fritzsch form. The FTY model is still consistent with experiments \cite{Fukugita:2003tn}. 
In the quark sector, the original Fritzsch quark mass matrices with six texture zeros are excluded \cite{Xing:2014sja}. while a straightforward extension of the Fritzsch ansatz
with four or five zeros are still allowed by current experimental data \cite{Xing:2015sva,Fritzsch:1999ee,Ibanez:1994ig,Ramond:1993kv,Xing:2003yj}.
In this sector, we will consider the RGE running behavior of quark mass matrices with four or five texture zeros and lepton mass matrices in FTY model.
\subsection{Four texture zeros}
In Hermitian basis, the quark mass matrices with four texture zeros are given by
\begin{eqnarray}
M^{}_{\rm u}\left(t^{}_{0}\right) = \begin{pmatrix} 
0 & C^{}_{\rm u} e^{ {\rm i} \rho^{}_{\rm u} } & 0 \\ 
C^{}_{\rm u} e^{- {\rm i} \rho^{}_{\rm u} } & \tilde B^{}_{\rm u} & B^{}_{\rm u} e^{ {\rm i} \sigma^{}_{\rm u} } \\ 
0 & B^{}_{\rm u} e^{- {\rm i} \sigma^{}_{\rm u} } & A^{}_{\rm u} 
\end{pmatrix} \;,\quad
M^{}_{\rm d}\left(t^{}_{0}\right) = \begin{pmatrix} 
0 & C^{}_{\rm d} e^{ {\rm i} \rho^{}_{\rm d} } & 0 \\ 
C^{}_{\rm d} e^{- {\rm i} \rho^{}_{\rm d} } & \tilde B^{}_{\rm d} & B^{}_{\rm d} e^{ {\rm i} \sigma^{}_{\rm d} } \\ 
0 & B^{}_{\rm d} e^{- {\rm i} \sigma^{}_{\rm d} } & A^{}_{\rm d} 
\end{pmatrix} \;.\label{42}
\end{eqnarray}
$M^{}_{\rm u}\left(t^{}_{0}\right)$ can be decomposed as 
\begin{eqnarray}
M^{}_{\rm u}\left(t^{}_{0}\right) = P \overline{M} P^\dagger \;, \quad {\rm with} \quad 
\overline{M} =  \begin{pmatrix} 
0 & C^{}_{\rm u} & 0 \\ 
C^{}_{\rm u} & \tilde B^{}_{\rm u} & B^{}_{\rm u} \\ 
0 & B^{}_{\rm u} & A^{}_{\rm u} 
\end{pmatrix} \;,\label{43}
\end{eqnarray}
and $P = {\rm Diag} \left\{1,e^{- {\rm i} \rho^{}_{\rm u} },e^{- {\rm i} \left( \rho^{}_{\rm u} + \sigma^{}_{\rm u} \right)} \right\} $.
The transformation matrix of  $M^{}_{\rm u}\left(t^{}_{0}\right)$ can be approximated as two successive rotations in $(2,3)$ and $(1,2)$ spaces,
\begin{eqnarray}
O^{}_{\rm u} = P \begin{pmatrix} c^{}_{\phi} & s^{}_{\phi} & 0 \\ - s^{}_{\phi} & c^{}_{\phi} & 0 \\ 0 & 0 & 1 \end{pmatrix} 
\begin{pmatrix} 1 & 0 & 0 \\ 0 & c^{}_{\theta} & s^{}_{\theta} \\ 0 & - s^{}_{\theta} & c^{}_{\theta} \end{pmatrix} \;,\label{44}
\end{eqnarray}
where $c^{}_{\phi}=\cos \phi$, $c^{}_{\theta}=\cos \theta$, $s^{}_{\phi}=\sin \phi$, $s^{}_{\theta}=\sin \theta$, 
$\phi \simeq C^{}_{\rm u} / (\tilde B^{}_{\rm u} - \frac{B^{2}_{\rm u}}{A^{}_{\rm u}} ) $ and $\theta \simeq B^{}_{\rm u} / (A^{}_{\rm u} - \tilde{B}^{}_{\rm u}) $. 
Because of the strong hierarchy of the up-type quark masses, the two angles $\phi$ and $\theta$ are considered as small quantities. 
Then the lead order contribution of RGE-corrected factors $ O^{}_{\rm u} I^{}_{\rm u,d} O^\dagger_{\rm u}$ in Eqs. (\ref{19}) and (\ref{25}) are given by
\begin{eqnarray}
O^{}_{\rm u} I^{}_{\rm u} O^\dagger_{\rm u} \simeq \begin{pmatrix}
1 & 0 & 0 \\ 
0 & 1 & \theta e^{ {\rm i} \sigma^{}_{\rm u} } \left( \xi^{3b/2}_{t} - 1 \right) \\ 
0 & \theta e^{- {\rm i} \sigma^{}_{\rm u} } \left( \xi^{3b/2}_{t} - 1 \right) & \xi^{3b/2}_{t}
\end{pmatrix} \;,\label{45}
\end{eqnarray} 
$ O^{}_{\rm u} I^{}_{\rm d} O^\dagger_{\rm u}$ is obtained by replacing $\xi^{3b/2}_{t}$ to $\xi^{3c/2}_{t}$. 
Interestingly, $\phi$ is irrelevant to the RGE running behavior.
With the help of Eqs. (\ref{19}) and  (\ref{25}), the RGE-corrected mass matrices are given by
%\begin{small}
\begin{eqnarray}
&&\hspace{-0.5cm} M^{}_{\rm u}\left(t\right) \simeq \gamma^{}_{\rm u} \begin{pmatrix}
0 & C^{}_{\rm u} e^{{\rm i} \rho^{}_{\rm u} } & 0 \\
C^{}_{\rm u} e^{- {\rm i} \rho^{}_{\rm u} } & \tilde B^{}_{\rm u} + \frac{B^{2}_{\rm u}}{A^{}_{\rm u}} \left( \xi^{3b/2}_{t} - 1 \right) &
x^{}_{1}%B^{}_{\rm u} e^{ {\rm i} \sigma^{}_{\rm u}} \left[ \xi^{3b/2}_{t} + \frac{\tilde B^{}_{\rm u}}{A^{}_{\rm u}} \left( \xi^{3b/2}_{t} - 1 \right) \right] 
\\
0 & x^{\dagger}_{1}%B^{}_{\rm u} e^{- {\rm i} \sigma^{}_{\rm u}} \left[ \xi^{3b/2}_{t} + \frac{\tilde B^{}_{\rm u}}{A^{}_{\rm u}} \left( \xi^{3b/2}_{t} - 1 \right) \right] 
& A^{}_{\rm u} \xi^{3b/2}_{t}
\end{pmatrix} \;,  \label{46}
\\ \nonumber \\ 
&&\hspace{-0.5cm} M^{}_{\rm d}\left(t\right) \simeq \gamma^{}_{\rm d} 
\begin{pmatrix}
0 & C^{}_{\rm d} e^{{\rm i} \rho^{}_{\rm d} } & 0 \\
C^{}_{\rm d} e^{- {\rm i} \rho^{}_{\rm d} } 
& \tilde B^{}_{\rm d} + \frac{B^{}_{\rm u}B^{}_{\rm d} }{A^{}_{\rm u}} e^{ {\rm i} \left( \sigma^{}_{\rm u} - \sigma^{}_{\rm d} \right) } \left( \xi^{3c/2}_{t} - 1 \right) &
x^{}_{2}%B^{}_{\rm d} e^{ {\rm i} \sigma^{}_{\rm d}}  
%\left[ 1 + \frac{A^{}_{\rm d} B^{}_{\rm u}}{A^{}_{\rm u} B^{}_{\rm d} } e^{ {\rm i} \left( \sigma^{}_{\rm u} - \sigma^{}_{\rm d} \right) } \left( \xi^{3c/2}_{t} - 1 \right) \right] 
\\ 0 &
x^{}_{3}%B^{}_{\rm d} e^{- {\rm i} \sigma^{}_{\rm d}} 
%\left[ \xi^{3c/2}_{t} + \frac{B^{}_{\rm u} \tilde B^{}_{\rm d} }{A^{}_{\rm u} B^{}_{\rm d}} e^{- {\rm i}  \left( \sigma^{}_{\rm u} - \sigma^{}_{\rm d} \right) } 
%\left( \xi^{3c/2}_{t} - 1 \right) \right] 
& x^{}_{4}%A^{}_{\rm d} \xi^{3c/2}_{t} + \frac{B^{}_{\rm u}B^{}_{\rm d} }{A^{}_{\rm u}} e^{- {\rm i} \left( \sigma^{}_{\rm u} - \sigma^{}_{\rm d} \right) } \left( \xi^{3c/2}_{t} - 1 \right) 
\end{pmatrix} \;, \label{47} 
\end{eqnarray}	
with 
\begin{eqnarray}
&& x^{}_{1} = B^{}_{\rm u} e^{ {\rm i} \sigma^{}_{\rm u}} \left[ \xi^{3b/2}_{t} + \frac{\tilde B^{}_{\rm u}}{A^{}_{\rm u}} \left( \xi^{3b/2}_{t} - 1 \right) \right] \;, \nonumber \\
&& x^{}_{2} = B^{}_{\rm d} e^{ {\rm i} \sigma^{}_{\rm d}}  
\left[ 1 + \frac{A^{}_{\rm d} B^{}_{\rm u}}{A^{}_{\rm u} B^{}_{\rm d} } e^{ {\rm i} \left( \sigma^{}_{\rm u} - \sigma^{}_{\rm d} \right) } \left( \xi^{3c/2}_{t} - 1 \right) \right] \;, \nonumber \\
&& x^{}_{3} = B^{}_{\rm d} e^{- {\rm i} \sigma^{}_{\rm d}} 
\left[ \xi^{3c/2}_{t} + \frac{B^{}_{\rm u} \tilde B^{}_{\rm d} }{A^{}_{\rm u} B^{}_{\rm d}} e^{- {\rm i}  \left( \sigma^{}_{\rm u} - \sigma^{}_{\rm d} \right) } 
\left( \xi^{3c/2}_{t} - 1 \right) \right]  \;, \nonumber \\
&& x^{}_{4} = A^{}_{\rm d} \xi^{3c/2}_{t} + \frac{B^{}_{\rm u}B^{}_{\rm d} }{A^{}_{\rm u}} e^{- {\rm i} \left( \sigma^{}_{\rm u} - \sigma^{}_{\rm d} \right) } \left( \xi^{3c/2}_{t} - 1 \right) \;, \nonumber 
\end{eqnarray}
%\end{small}
where we have neglected the terms which are proportional to $ \theta C^{}_{\rm u,d}$.
In this approximation, the four texture zeros of $M^{}_{\rm u}$ and $M^{}_{\rm d}$ are stable against the RGE effects.
Eq. (\ref{46}) and (\ref{47}) are similar to the result of Ref. \cite{Xing:2015sva} where a different approximation is adopted.
As described in the previous section, $M^{}_{\rm u}\left(t\right)$ remains Hermitian and $M^{}_{\rm d}\left(t\right)$ becomes non-Hermitian. 
One can transform $M^{}_{\rm d}\left(t\right)$ to be Hermitian by rotating the right-handed down-type quark fields, 
the transformed $M'^{}_{\rm d}\left(t\right)$ has a very complicated form with no texture zero, then we do not show it.
Furthermore, in Eq. (\ref{47}) only the phase different $\sigma^{}_{\rm u} - \sigma^{}_{\rm d}$ rather than the phases themselves is relevant to the RGE-corrections and can change
the mass matrix $M^{}_{\rm d}$ in a non-trivial way.
\subsection{Five texture zeros}
Ramond, Roberts and Ross (RRR) have found that there exist five phenomenologically allowed patterns of Hermitian quark mass matrices with five texture zeros \cite{Ramond:1993kv}.
The magnitudes of the elements of five RRR-type quark mass matrices are approximately given in Ref. \cite{Fritzsch:1999ee}. Based on these result, we can obtain the approximate form 
$O^{}_{\rm u}$ and RGE corrected quark mass matrices.

 $\bullet$ Pattern I
\begin{eqnarray}
M^{}_{\rm u} \left(t^{}_{0}\right) = \begin{pmatrix}
0 & C^{}_{\rm u} e^{ {\rm i} \rho^{}_{\rm u} } & 0 \\ 
C^{}_{\rm u} e^{- {\rm i} \rho^{}_{\rm u} } & \tilde B^{}_{\rm u} & 0 \\ 
0 & 0 & A^{}_{\rm u} 
\end{pmatrix} \;,\quad 
M^{}_{\rm d} \left(t^{}_{0}\right) = \begin{pmatrix} 
0 & C^{}_{\rm d} e^{ {\rm i} \rho^{}_{\rm d} } & 0 \\ 
C^{}_{\rm d} e^{- {\rm i} \rho^{}_{\rm d} } & \tilde B^{}_{\rm d} & B^{}_{\rm d} e^{ {\rm i} \sigma^{}_{\rm d} } \\ 
0 & B^{}_{\rm d} e^{- {\rm i} \sigma^{}_{\rm d} } & A^{}_{\rm d} 
\end{pmatrix} \;.
\end{eqnarray}
The transformation matrix of $M^{}_{\rm u} \left(t^{}_{0}\right) $ is
\begin{eqnarray}
O^{}_{\rm u} = \begin{pmatrix} 
c^{}_{\theta} & s^{}_{\theta} & 0 \\ 
- s^{}_{\theta} e^{- {\rm i} \rho^{}_{\rm u} } &  c^{}_{\theta} e^{- {\rm i} \rho^{}_{\rm u} } & 0 \\ 
0   &   0   &  1
\end{pmatrix} \;,
\end{eqnarray}
with $\theta \simeq C^{}_{\rm u} / \tilde B^{}_{\rm u}$. The RGE corrected quark mass matrices are given by
\begin{eqnarray}
\hspace{-1cm} M^{}_{\rm u}\left(t\right) \simeq \gamma^{}_{\rm u}
\begin{pmatrix}
0 & C^{}_{\rm u} e^{ {\rm i} \rho^{}_{\rm u} } & 0 \\ 
C^{}_{\rm u} e^{- {\rm i} \rho^{}_{\rm u} } & \tilde B^{}_{\rm u} & 0 \\ 
0 & 0 & A^{}_{\rm u} \xi^{3b/2}_{t}
\end{pmatrix} \;,\quad
M^{}_{\rm d}\left(t\right) \simeq \gamma^{}_{\rm d} 
\begin{pmatrix} 
0 & C^{}_{\rm d} e^{ {\rm i} \rho^{}_{\rm d} } & 0 \\ 
C^{}_{\rm d} e^{- {\rm i} \rho^{}_{\rm d} } & \tilde B^{}_{\rm d} & B^{}_{\rm d} e^{ {\rm i} \sigma^{}_{\rm d} } \\ 
0 & B^{}_{\rm d} \xi^{3c/2}_{t} e^{- {\rm i} \sigma^{}_{\rm d} } & A^{}_{\rm d} \xi^{3c/2}_{t}
\end{pmatrix} \;.
\end{eqnarray}

 $\bullet$ Pattern II
\begin{eqnarray}
M^{}_{\rm u} \left(t^{}_{0}\right) = \begin{pmatrix}
0 & C^{}_{\rm u} e^{ {\rm i} \rho^{}_{\rm u} } & 0 \\ 
C^{}_{\rm u} e^{- {\rm i} \rho^{}_{\rm u} } & 0 & B^{}_{\rm u} e^{ {\rm i} \sigma^{}_{\rm u} } \\ 
0 & B^{}_{\rm u} e^{- {\rm i} \sigma^{}_{\rm u} } & A^{}_{\rm u} 
\end{pmatrix} \;,\quad 
M^{}_{\rm d} \left(t^{}_{0}\right) = \begin{pmatrix} 
0 & C^{}_{\rm d} e^{ {\rm i} \rho^{}_{\rm d} } & 0 \\ 
C^{}_{\rm d} e^{- {\rm i} \rho^{}_{\rm d} } & \tilde B^{}_{\rm d} & B^{}_{\rm d} e^{ {\rm i} \sigma^{}_{\rm d} } \\ 
0 & B^{}_{\rm d} e^{- {\rm i} \sigma^{}_{\rm d} } & A^{}_{\rm d} 
\end{pmatrix} \;,
\end{eqnarray}
The transformation matrix of $M^{}_{\rm u} \left(t^{}_{0}\right) $ is
\begin{eqnarray}
O^{}_{\rm u} = P \begin{pmatrix} c^{}_{\phi} & s^{}_{\phi} & 0 \\ - s^{}_{\phi} & c^{}_{\phi} & 0 \\ 
0 & 0 & 1 \end{pmatrix}
\begin{pmatrix} 1 & 0 & 0 \\ 0 & c^{}_{\theta} & s^{}_{\theta} \\ 0 & - s^{}_{\theta} & c^{}_{\theta} \end{pmatrix} \;,
\end{eqnarray}
where $P = {\rm Diag} \left\{ 1 , e^{- {\rm i} \rho^{}_{u} } , e^{ - {\rm i} (\rho^{}_{\rm u} + \sigma^{}_{\rm u}) } \right\}$,
$\theta \simeq B^{}_{\rm u} / A^{}_{\rm u} $ and $\phi \simeq (A^{}_{\rm u} C^{}_{\rm u})/B^{}_{\rm u}$. 
The RGE corrected quark mass matrices are given by
\begin{eqnarray}
&&\hspace{-1cm} M^{}_{\rm u}\left(t\right) \simeq \gamma^{}_{\rm u} 
\begin{pmatrix}
0 & C^{}_{\rm u} e^{ {\rm i} \rho^{}_{\rm u} } & 0 \\ 
C^{}_{\rm u} e^{- {\rm i} \rho^{}_{\rm u} } & \frac{B^2_{\rm u}}{A^{}_{\rm u}} \left( \xi^{3b/2}_{t} -1 \right) & B^{}_{\rm u} \xi^{3b/2}_{t} e^{ {\rm i} \sigma^{}_{\rm u} } \\ 
0 & B^{}_{\rm u} \xi^{3b/2}_{t} e^{- {\rm i} \sigma^{}_{\rm u} } & A^{}_{\rm u} \xi^{3b/2}_{t}
\end{pmatrix} \;, \nonumber \\
&&\hspace{-1cm} M^{}_{\rm d}\left(t\right) \simeq \gamma^{}_{\rm d} 
\begin{pmatrix}
0 & C^{}_{\rm d} e^{{\rm i} \rho^{}_{\rm d} } & 0 \\
C^{}_{\rm d} e^{- {\rm i} \rho^{}_{\rm d} } 
& \tilde B^{}_{\rm d} + \frac{B^{}_{\rm u} B^{}_{\rm d} }{A^{}_{\rm u}} \left( \xi^{3c/2}_{t} - 1 \right) e^{ {\rm i} \left( \sigma^{}_{\rm u} - \sigma^{}_{\rm d} \right) } &
B^{}_{\rm d} e^{ {\rm i} \sigma^{}_{\rm d}} + \frac{A^{}_{\rm d} B^{}_{\rm u}}{A^{}_{\rm u}} \left( \xi^{3c/2}_{t} - 1 \right) e^{ {\rm i} \sigma^{}_{\rm u}} \\
0 &
B^{}_{\rm d} \xi^{3c/2}_{t} e^{- {\rm i} \sigma^{}_{\rm d}} + \frac{\tilde B^{}_{\rm d} B^{}_{\rm u}}{A^{}_{\rm u}} \left( \xi^{3c/2}_{t} - 1 \right) e^{- {\rm i} \sigma^{}_{\rm u}} &
A^{}_{\rm d} \xi^{3c/2}_{t} %+ \frac{B^{}_{\rm u} B^{}_{\rm d} }{A^{}_{\rm u}} \left( \xi^{3c/2}_{t} - 1 \right) e^{- {\rm i} \left( \sigma^{}_{\rm u} - \sigma^{}_{\rm d} \right) } 
\end{pmatrix} \;.\nonumber \\
\end{eqnarray}
 $\bullet$ Pattern III
\begin{eqnarray}
M^{}_{\rm u} \left(t^{}_{0}\right) = \begin{pmatrix}
0 & 0 & F^{}_{\rm u} e^{ {\rm i} \eta^{}_{\rm u} } \\ 
0 & \tilde B^{}_{\rm u} & 0 \\ 
F^{}_{\rm u} e^{- {\rm i} \eta^{}_{\rm u} } & 0 & A^{}_{\rm u} 
\end{pmatrix} \;,\quad 
M^{}_{\rm d} \left(t^{}_{0}\right) = \begin{pmatrix} 
0 & C^{}_{\rm d} e^{ {\rm i} \rho^{}_{\rm d} } & 0 \\ 
C^{}_{\rm d} e^{- {\rm i} \rho^{}_{\rm d} } & \tilde B^{}_{\rm d} & B^{}_{\rm d} e^{ {\rm i} \sigma^{}_{\rm d} } \\ 
0 & B^{}_{\rm d} e^{- {\rm i} \sigma^{}_{\rm d} } & A^{}_{\rm d} 
\end{pmatrix} \;,
\end{eqnarray}
The transformation matrix of $M^{}_{\rm u} \left(t^{}_{0}\right) $ is
\begin{eqnarray}
O^{}_{\rm u} = \begin{pmatrix} 
c^{}_{\theta} & 0 & s^{}_{\theta} \\ 
0 & e^{- {\rm i} \eta^{}_{\rm u} } & 0 \\ 
- s^{}_{\theta} e^{- {\rm i} \eta^{}_{\rm u} } & 0 & c^{}_{\theta} e^{- {\rm i} \eta^{}_{\rm u} }
\end{pmatrix} \;,
\end{eqnarray}
with $ \theta \simeq F^{}_{\rm u} / A^{}_{\rm u} $. The RGE corrected quark mass matrices are given by
\begin{eqnarray}
&&\hspace{-1cm} M^{}_{\rm u}\left(t\right) \simeq \gamma^{}_{\rm u} 
\begin{pmatrix}
0 & 0 & F^{}_{\rm u} \xi^{3b/2}_{t} e^{ {\rm i} \eta^{}_{\rm u} } \\ 
0 & \tilde B^{}_{\rm u} & 0 \\ 
F^{}_{\rm u} \xi^{3b/2}_{t} e^{- {\rm i} \eta^{}_{\rm u} } & 0 & A^{}_{\rm u} \xi^{3b/2}_{t}
\end{pmatrix} \;, \nonumber \\
&&\hspace{-1cm} M^{}_{\rm d}\left(t\right) \simeq \gamma^{}_{\rm d}  
\begin{pmatrix} 
0 & C^{}_{\rm d} e^{ {\rm i} \rho^{}_{\rm d} } 
+ \frac{B^{}_{\rm d} F^{}_{\rm u}}{A^{}_{\rm u}} \left( \xi^{3c/2}_{t} - 1 \right) e^{ {\rm i} \left( \eta^{}_{\rm u} - \sigma^{}_{\rm d} \right) } 
& \frac{A^{}_{\rm d} F^{}_{\rm u}}{A^{}_{\rm u}} \left( \xi^{3c/2}_{t} - 1 \right) e^{ {\rm i} \eta^{}_{\rm u} } \\ 
C^{}_{\rm d} e^{- {\rm i} \rho^{}_{\rm d} } & \tilde B^{}_{\rm d} & B^{}_{\rm d} e^{ {\rm i} \sigma^{}_{\rm d} } \\ 
0 & B^{}_{\rm d} \xi^{3c/2}_{t} e^{- {\rm i} \sigma^{}_{\rm d} } & A^{}_{\rm d} \xi^{3c/2}_{t} 
\end{pmatrix} \;.
\end{eqnarray}

 $\bullet$ Pattern IV
\begin{eqnarray}
M^{}_{\rm u} \left(t^{}_{0}\right) = \begin{pmatrix}
0 & C^{}_{\rm u} e^{ {\rm i} \rho^{}_{\rm u} } & 0 \\ 
C^{}_{\rm u} e^{- {\rm i} \rho^{}_{\rm u} } & \tilde B^{}_{\rm u} & B^{}_{\rm u} e^{ {\rm i} \sigma^{}_{\rm u} } \\ 
0 & B^{}_{\rm u} e^{- {\rm i} \sigma^{}_{\rm u} } & A^{}_{\rm u} 
\end{pmatrix} \;,\quad 
M^{}_{\rm d} \left(t^{}_{0}\right) = \begin{pmatrix} 
0 & C^{}_{\rm d} e^{ {\rm i} \rho^{}_{\rm d} } & 0 \\ 
C^{}_{\rm d} e^{- {\rm i} \rho^{}_{\rm d} } & \tilde B^{}_{\rm d} & 0 \\ 
0 & 0 & A^{}_{\rm d} 
\end{pmatrix} \;,
\end{eqnarray}
The transformation matrix of $M^{}_{\rm u} \left(t^{}_{0}\right) $ is
\begin{eqnarray}
O^{}_{\rm u} = P \begin{pmatrix} c^{}_{\phi} & s^{}_{\phi} & 0 \\ - s^{}_{\phi} & c^{}_{\phi} & 0 \\ 
0 & 0 & 1 \end{pmatrix}
\begin{pmatrix} 1 & 0 & 0 \\ 0 & c^{}_{\theta} & s^{}_{\theta} \\ 0 & - s^{}_{\theta} & c^{}_{\theta} \end{pmatrix} \;,
\end{eqnarray}
where $P = {\rm Diag} \left\{ 1 , e^{- {\rm i} \rho^{}_{u} } , e^{ - {\rm i} (\rho^{}_{\rm u} + \sigma^{}_{\rm u}) } \right\}$,
$\theta \simeq B^{}_{\rm u} / (A^{}_{\rm u} - \tilde B^{}_{\rm u}) $ and $\phi \simeq C^{}_{\rm u} / (\tilde B^{}_{\rm u} - \frac{B^{2}_{\rm u}}{A^{}_{\rm u}} ) $.
The RGE corrected quark mass matrices are given by
\begin{eqnarray}
&&\hspace{-0.5cm} M^{}_{\rm u}\left(t\right) \simeq \gamma^{}_{\rm u}
\begin{pmatrix}
0 & C^{}_{\rm u} e^{{\rm i} \rho^{}_{\rm u} } & 0 \\
C^{}_{\rm u} e^{- {\rm i} \rho^{}_{\rm u} } & \tilde B^{}_{\rm u} + \frac{B^{2}_{\rm u}}{A^{}_{\rm u}} \left( \xi^{3b/2}_{t} - 1 \right) &
B^{}_{\rm u} e^{ {\rm i} \sigma^{}_{\rm u}} \left[ \xi^{3b/2}_{t} + \frac{\tilde B^{}_{\rm u}}{A^{}_{\rm u}} \left( \xi^{3b/2}_{t} - 1 \right) \right] \\
0 & B^{}_{\rm u} e^{- {\rm i} \sigma^{}_{\rm u}} \left[ \xi^{3b/2}_{t} + \frac{\tilde B^{}_{\rm u}}{A^{}_{\rm u}} \left( \xi^{3b/2}_{t} - 1 \right) \right] &
A^{}_{\rm u} \xi^{3b/2}_{t}
\end{pmatrix} \;, \nonumber
\\ \nonumber \\ 
&&\hspace{-0.5cm} M^{}_{\rm d}\left(t\right) \simeq \gamma^{}_{\rm d} 
\begin{pmatrix} 
0 & C^{}_{\rm d} e^{ {\rm i} \rho^{}_{\rm d} } & 0 \\ 
C^{}_{\rm d} e^{- {\rm i} \rho^{}_{\rm d} } & \tilde B^{}_{\rm d} & \frac{ A^{}_{\rm d} B^{}_{\rm u} }{A^{}_{\rm u}} \left( \xi^{3c/2}_{t} - 1 \right) e^{ {\rm i} \sigma^{}_{\rm u} } \\ 
0 & \frac{ B^{}_{\rm u} \tilde B^{}_{\rm d} }{A^{}_{\rm u}} \left( \xi^{3c/2}_{t} - 1 \right) e^{- {\rm i} \sigma^{}_{\rm u} }  & A^{}_{\rm d} \xi^{3b/2}_{t}
\end{pmatrix} \;. 
\end{eqnarray}
 $\bullet$ Pattern V
\begin{eqnarray}
M^{}_{\rm u} \left(t^{}_{0}\right) = \begin{pmatrix}
0 & 0 & F^{}_{\rm u} e^{ {\rm i} \eta^{}_{\rm u} } \\ 
0 & \tilde B^{}_{\rm u} & B^{}_{\rm u} e^{ {\rm i} \sigma^{}_{\rm u} } \\ 
F^{}_{\rm u} e^{- {\rm i} \eta^{}_{\rm u} } & B^{}_{\rm u} e^{- {\rm i} \sigma^{}_{\rm u} } & A^{}_{\rm u} 
\end{pmatrix} \;,\quad 
M^{}_{\rm d} \left(t^{}_{0}\right) = \begin{pmatrix} 
0 & C^{}_{\rm d} e^{ {\rm i} \rho^{}_{\rm d} } & 0 \\ 
C^{}_{\rm d} e^{- {\rm i} \rho^{}_{\rm d} } & \tilde B^{}_{\rm d} & 0 \\ 
0 & 0 & A^{}_{\rm d} 
\end{pmatrix} \;,
\end{eqnarray}
The transformation matrix of $M^{}_{\rm u} \left(t^{}_{0}\right) $ is
\begin{eqnarray}
O^{}_{\rm u} = P \begin{pmatrix} c^{}_{\phi} & s^{}_{\phi} & 0 \\ - s^{}_{\phi} & c^{}_{\phi} & 0 \\ 0 & 0 & 1 \end{pmatrix}
\begin{pmatrix} c^{}_{\alpha} & 0 & s^{}_{\alpha} \\ 0 & 1 & 0 \\ - s^{}_{\alpha} & 0 & c^{}_{\alpha} \end{pmatrix}
\begin{pmatrix} 1 & 0 & 0 \\ 0 & c^{}_{\theta} & s^{}_{\theta} \\ 0 & - s^{}_{\theta} & c^{}_{\theta} \end{pmatrix} \;,
\end{eqnarray}
with $P = {\rm Diag} \left\{ e^{- {\rm i} \sigma^{}_{u} } , e^{- {\rm i} \eta^{}_{u} } , e^{ - {\rm i} (\eta^{}_{\rm u} + \sigma^{}_{\rm u}) } \right\}$,
$\theta \simeq B^{}_{\rm u} / \left( A - \tilde{B^{}_{\rm u}} \right)$, $\alpha \simeq F^{}_{\rm u} / A^{}_{\rm u} $ and $\phi \simeq \theta F^{}_{\rm u} / \tilde{B^{}_{\rm u}} $.
The RGE corrected quark mass matrices are given by
\begin{eqnarray}
&& M^{}_{\rm u} \left(t\right) \simeq \begin{pmatrix}
0 & 0 & F^{}_{\rm u} \xi^{3b/2}_{t} e^{ {\rm i} \eta^{}_{\rm u} } \\ 
0 & \tilde B^{}_{\rm u} + \frac{B^{2}_{\rm u}}{A^{}_{\rm u}} \left( \xi^{3b/2}_{t} - 1 \right)  
& B^{}_{\rm u} e^{ {\rm i} \sigma^{}_{\rm u}} \left[ \xi^{3b/2}_{t} + \frac{\tilde B^{}_{\rm u}}{A^{}_{\rm u}} \left( \xi^{3b/2}_{t} - 1 \right) \right] \\ 
F^{}_{\rm u} \xi^{3b/2}_{t} e^{- {\rm i} \eta^{}_{\rm u} } 
& B^{}_{\rm u} e^{- {\rm i} \sigma^{}_{\rm u}} \left[ \xi^{3b/2}_{t} + \frac{\tilde B^{}_{\rm u}}{A^{}_{\rm u}} \left( \xi^{3b/2}_{t} - 1 \right) \right] 
& A^{}_{\rm u} \xi^{3b/2}_{t} 
\end{pmatrix} \;, \nonumber \\
&& M^{}_{\rm d} \left(t\right) \simeq \begin{pmatrix} 
0 & C^{}_{\rm d} e^{ {\rm i} \rho^{}_{\rm d} } & \frac{A^{}_{\rm d} F^{}_{\rm u}}{A^{}_{\rm u}} e^{ {\rm i} \eta^{}_{\rm u} } \left( \xi^{3c/2}_{t} - 1 \right) \\ 
C^{}_{\rm d} e^{- {\rm i} \rho^{}_{\rm d} } & \tilde B^{}_{\rm d} 
& \frac{A^{}_{\rm d} B^{}_{\rm u}}{A^{}_{\rm u}} e^{ {\rm i} \rho^{}_{\rm u} } \left( \xi^{3c/2}_{t} - 1 \right) \\ 
\frac{B^{}_{\rm u} C^{}_{\rm d}}{A^{}_{\rm u}} e^{- {\rm i} \left( \sigma^{}_{\rm u} + \rho^{}_{\rm d} \right) } \left( \xi^{3c/2}_{t} - 1 \right) 
& \frac{B^{}_{\rm u} \tilde{B}^{}_{\rm d}}{A^{}_{\rm u}} e^{- {\rm i} \sigma^{}_{\rm u} } \left( \xi^{3c/2}_{t} - 1 \right) 
& A^{}_{\rm d} \xi^{3c/2}_{t} 
\end{pmatrix} \;.
\end{eqnarray}

In some reliable approximations, $M^{}_{\rm u}$ and $M^{}_{\rm d}$ of pattern I, $M^{}_{\rm d}$ of pattern II.
$M^{}_{\rm u}$ of pattern III, IV and V, are stable against the RGE effects.
\subsection{FTY model}
The charged lepton mass matrix $M^{}_{\ell}$ and Dirac mass matrix $M^{}_{\nu D}$ in the FTY model have the following Fritzsch structures \cite{Fukugita:1992sy} 
\begin{eqnarray}
M^{}_{\nu D} = \begin{pmatrix} 
0 & F^{}_{\nu} e^{ {\rm i} \rho^{}_{\nu} } & 0 \\ 
F^{}_{\nu} e^{- {\rm i} \rho^{}_{\nu} } & 0 & B^{}_{\nu} e^{ {\rm i} \sigma^{}_{\nu} } \\ 
0 & B^{}_{\nu} e^{- {\rm i} \sigma^{}_{\nu} } & A^{}_{\nu} 
\end{pmatrix} \;,\quad
M^{}_{\ell} = \begin{pmatrix} 
0 & F^{}_{\ell} e^{ {\rm i} \rho^{}_{\ell} } & 0 \\ 
F^{}_{\ell} e^{- {\rm i} \rho^{}_{\ell} } & 0 & B^{}_{\ell} e^{ {\rm i} \sigma^{}_{\ell} } \\ 
0 & B^{}_{\ell} e^{- {\rm i} \sigma^{}_{\ell} } & A^{}_{\ell} 
\end{pmatrix} \;.
\end{eqnarray}
For simplicity, the Majorana mass of the right-handed neutrino $M^{}_{R}$ is taken to be proportional to the unit matrix in the basis in which $M^{}_{\nu D}$ is diagonal.
The effective light neutrino mass matrix is given by the seesaw relation
\begin{eqnarray}
M^{}_{\nu} = \frac{1}{M^{}_{R} } M^{}_{\nu D} M^{T}_{\nu D} \;.
\end{eqnarray}
Similar to the case of four texture zeros, 
for $m^{}_{\nu^{}_{e}} \ll m^{}_{\nu^{}_{\mu}} \ll m^{}_{\nu^{}_{\tau}} $ and $m^{}_{e} \ll m^{}_{\mu} \ll m^{}_{\tau} $,  the transformation matrices $O^{}_{\rm f}$ can be parameterized as 
\begin{eqnarray}
O^{}_{\rm f} = P \begin{pmatrix} \cos \phi^{}_{\rm f} & \sin \phi^{}_{\rm f} & 0 \\ - \sin \phi^{}_{\rm f} & \cos \phi^{}_{\rm f} & 0 \\ 
0 & 0 & 1 \end{pmatrix}
\begin{pmatrix} 1 & 0 & 0 \\ 0 & \cos \theta^{}_{\rm f} & \sin \theta^{}_{\rm f} \\ 0 & - \sin \theta^{}_{\rm f} & \cos \theta^{}_{\rm f} \end{pmatrix} \;,
\end{eqnarray}
where ${\rm f} = \ell$ and $\nu$ and
\begin{eqnarray}
&& P^{}_{f} = {\rm Diag} \left\{ 1 , e^{- {\rm i} \rho^{}_{f} } , e^{ - {\rm i} (\rho^{}_{f} + \sigma^{}_{f}) } \right\} \;,\nonumber \\
&& \phi^{}_{\ell} \simeq \frac{ A^{}_{\ell} F^{}_{\ell} }{ B^2_{\ell} } \simeq \sqrt{ \frac{ m^{}_{e} }{ m^{}_{\mu} } } \;,\qquad 
\theta^{}_{\ell} \simeq \frac{ B^{}_{\ell} }{ A^{}_{\ell} } \simeq \sqrt{ \frac{ m^{}_{\mu} }{ m^{}_{\tau} } } \;. \nonumber \\
&& \phi^{}_{\nu} \simeq \frac{ A^{}_{\nu} F^{}_{\nu} }{ B^2_{\nu} } \simeq \sqrt[4]{ \frac{ m^{}_{\nu^{}_{e}} }{ m^{}_{\nu^{}_{\mu}} } } \;,\quad 
\theta^{}_{\nu} \simeq \frac{ B^{}_{\nu} }{ A^{}_{\nu} } \simeq \sqrt[4]{ \frac{ m^{}_{\nu^{}_{\mu}} }{ m^{}_{\nu^{}_{\tau}} } } \;.
\end{eqnarray}
The RGE-corrected lepton mass matrices are given by
\begin{eqnarray}
M^{}_{l}\left(t\right) &=& \gamma^{}_{l} O^{}_{l} I^{}_{l} O^{\dagger}_{l} M^{}_{l} \left(t^{}_{0}\right) 
= \gamma^{}_{l} \begin{pmatrix} 
0 & F^{}_{\ell} e^{ {\rm i} \rho^{}_{\ell} } & 0 \\ 
F^{}_{\ell} e^{- {\rm i} \rho^{}_{\ell} } & \frac{B^{2}_{\ell} }{A^{}_{\ell} } \left( \xi^{3b/2}_{\tau} - 1 \right) 
& B^{}_{\ell} \xi^{3b/2}_{\tau} e^{ {\rm i} \sigma^{}_{\ell} } \\ 
0 & B^{}_{\ell} \xi^{3b/2}_{\tau} e^{- {\rm i} \sigma^{}_{\ell} }  & A^{}_{\ell} \xi^{3b/2}_{\tau}
\end{pmatrix} \;, \nonumber \\
M^{}_{\nu}\left(t\right) &=& \frac{\gamma^{}_{\nu} }{M^{}_{R} } A A^T \;,\nonumber \\
A &=& O^{}_{l} I^{}_{\nu} O^{\dagger}_{l} M^{}_{\nu D} \left(t^{}_{0}\right) \nonumber \\
&=& \begin{pmatrix}
0 & F^{}_{\nu} e^{ {\rm i} \rho^{}_{\nu} } & 0 \\
F^{}_{\nu} e^{- {\rm i} \rho^{}_{\nu} } & \frac{B^{}_{\ell} B^{}_{\nu} }{A^{}_{\ell} } \left( \xi^{C^{}_{\nu}}_{\tau} -1 \right) e^{ {\rm i} \left( \sigma^{}_{\ell} - \sigma^{}_{\nu} \right) }
& B^{}_{\nu} e^{ {\rm i} \sigma^{}_{\nu} } + \frac{A^{}_{\nu} B^{}_{\ell} }{A^{}_{\ell} } \left( \xi^{C^{}_{\nu}}_{\tau} -1 \right) e^{ {\rm i} \rho^{}_{\ell} } \\
0 & B^{}_{\nu} e^{- {\rm i} \sigma^{}_{\nu} } \xi^{C^{}_{\nu}}_{\tau} 
& A^{}_{\nu} \xi^{C^{}_{\nu}}_{\tau} + \frac{B^{}_{\ell} B^{}_{\nu}}{A^{}_{\ell}} \left( \xi^{C^{}_{\nu}}_{\tau} - 1 \right) e^{- {\rm i} \left( \sigma^{}_{\ell} - \sigma^{}_{\nu} \right) }
\end{pmatrix} \;.
\end{eqnarray}
Obviously, the charged lepton obtain non-zero 22 element. If we neglect the terms which are proportional to $ \theta^{}_{l} F^{}_{\nu}$, both of $M^{}_{\nu}\left(t\right)$ and 
$M^{}_{\nu}\left(t^{}_{0}\right)$ are of the from
\begin{eqnarray}
\begin{pmatrix}
\times &   0    & \times \\
0      & \times & \times \\
\times & \times & \times
\end{pmatrix} \;,
\end{eqnarray}
where $\times$ denotes non-zero elements.

\section{Tribimaximal mixing}
In this section we consider the RGE effects of the lepton mass matrix with a tribimaximal mixing pattern. Based on the non–Abelian discrete symmetry $A^{}_{4}$, Ref. \cite{Babu:2005se} has present a renormalizable gauge models which realize such mixing pattern naturally. 
In that model, the light charged leptons and neutrino mass matrices are given by
\begin{eqnarray}
M^{}_{l} \left(t^{}_{0}\right) = O^{}_{l} \begin{pmatrix}
m^{}_{e} & & \\ & m^{}_{\mu} & \\  & & m^{}_{\tau}
\end{pmatrix} \;,\quad
M^{}_{\nu} \left(t^{}_{0}\right) = m^{}_{0} \begin{pmatrix}
1 & 0 & x \\ -0 & 1 - x^2 & 0 \\ x & 0 & 1
\end{pmatrix} \;,
\end{eqnarray}
where $m^{}_{0}$ is real, the complex parameter $x$ can be defined as $x = |x|e^{ {\rm i} \psi } $ , and  
\begin{eqnarray}
O^{}_{l} = \frac{1}{\sqrt 3} \begin{pmatrix}
1 & 1 & 1 \\ 1 & \omega & \omega^2 \\ 1 & \omega^2 & \omega 
\end{pmatrix} \;,
\end{eqnarray}
with $\omega = e^{ 2 {\rm i} \pi /3 }$. $M^{}_{\nu}$ can be diagonalized by the transformation $M^{}_{\nu} = O^{}_{\nu} D^{}_{\nu} O^{T}_{\nu} $ with
\begin{eqnarray}
O^{}_{\nu} = \frac{1}{\sqrt 2} \begin{pmatrix}
1 & 0 & -1 \\ 0 & \sqrt 2 & 0 \\ 1 & 0 & 1
\end{pmatrix} P \;, \quad
D^{}_{\nu} = m^{}_{0} \begin{pmatrix}
|1+x| & & \\ & |1-x^2| & \\  & & |1-x|
\end{pmatrix}  \;,
\end{eqnarray}
where $P$ is a diagonal phase matrix. The PMNS matrix is given by 
\begin{eqnarray}
O^{}_{\rm PMNS} = O^{\dagger}_{l} O^{}_{\nu} = P' \begin{pmatrix}
\sqrt{ \frac{2}{3} } & \frac{1}{\sqrt 3} & 0 \\
- \frac{1}{\sqrt 6}  & \frac{1}{\sqrt 3} & - \frac{1}{\sqrt 2} \\
- \frac{1}{\sqrt 6}  & \frac{1}{\sqrt 3} & \frac{1}{\sqrt 2}
\end{pmatrix} P \;, \label{TB}
\end{eqnarray}
where $P'$ is the nonphysical phase matrix.

According to Eq.~\ref{39}, the RGE-corrected lepton mass matrices are given by
\begin{eqnarray}
M^{}_{l}\left(t\right) &=& \gamma^{}_{l} O^{}_{l} I^{}_{l} O^{\dagger}_{l} M^{}_{l} \left(t^{}_{0}\right)
= \gamma^{}_{l} O^{}_{l} I^{}_{l} \begin{pmatrix}
m^{}_{e} & 0 & 0 \\0 & m^{}_{\mu} & 0 \\ 0 & 0 & m^{}_{\tau}
\end{pmatrix} \;,
\\ \nonumber \\
M^{}_{\nu}\left(t\right) &=& \gamma^{}_{\nu} O^{}_{l} I^{}_{\nu} O^{\dagger}_{l} M^{}_{\nu} \left(t^{}_{0}\right) O^{*}_{l} I^{}_{\nu} O^{T}_{l} % \nonumber\
\simeq \gamma^{}_{\nu} M^{}_{\nu} \left(t^{}_{0}\right) + \frac{\gamma^{}_{\nu} \epsilon }{3} M^{}_{1} \;,
\end{eqnarray}
where the first order of $\epsilon$ is kept, $\epsilon$ and $M^{}_{1}$ are defined as 
\begin{eqnarray}
\epsilon = \xi^{C^{}_{\nu}}_{\tau}-1 \;,\quad
M^{}_{1} = m^{}_{0} \begin{pmatrix}
2 x \omega ^2+2 & x^2 (-\omega )+x \omega -1 & 2 x-1 \\
x^2 (-\omega )+x \omega -1 & 2-2 x^2 & -x^2 \omega ^2+x \omega ^2-1 \\
2 x-1 & -x^2 \omega ^2+x \omega ^2-1 & 2 x \omega +2
\end{pmatrix} \;.
\end{eqnarray}
Since $O^{}_{l}$ is independent of scale, we only need to diagonalize $M^{}_{\nu}\left(t\right)$. Firstly, we define the Hermitian matrix 
\begin{eqnarray}
H^{}_{\nu}\left(t\right) &=& M^{}_{\nu}\left(t\right) M^{\dagger}_{\nu}\left(t\right) \nonumber \\
&=& \gamma^{2}_{\nu} M^{}_{\nu} \left(t^{}_{0}\right) M^{*}_{\nu} \left(t^{}_{0}\right) 
+ \frac{\gamma^{2}_{\nu} \epsilon}{3}~ \Big[ M^{}_{\nu} \left(t^{}_{0}\right) M^{*}_{1} + M^{}_{1} M^{}_{\nu} \left(t^{}_{0}\right) \Big] 
+ \frac{\gamma^{2}_{\nu} \epsilon^2}{9} M^{}_{1} M^{*}_{1} \;.
\end{eqnarray}
The eigenvalues are 
\begin{eqnarray}
&& m^{2}_{1} \simeq m^{2}_{0} \left( 1 + \frac{2}{3} \epsilon \right) \left( 1 + 2 |x| c^{}_{\phi} + |x|^2 \right) \;,\quad \nonumber \\
&& m^{2}_{2} \simeq m^{2}_{0} \left( 1 + \frac{4}{3} \epsilon \right) \left( 1 - 2 |x|^2 \cos 2 \phi + |x|^4 \right) \;,\quad \nonumber \\
&& m^{2}_{3} \simeq m^{2}_{0} \left( 1 + 2 \epsilon \right) \left( 1 - 2 |x| c^{}_{\phi} + |x|^2 \right) \;. \label{80}
\end{eqnarray}
The normalized eigenvectors are
\begin{eqnarray}
 v^{}_{1} &\simeq& \Big\{ \frac{1}{\sqrt 2} + \frac{\epsilon \left( s^{}_{\phi} - 2 {\rm i} |x| \right)}{2 c^{}_\phi \sqrt 6}  \;,\;
\frac{ \epsilon~\left( |x|^2 - 2 {\rm i} |x| s^{}_{\phi} - 4 |x| c^{}_{\phi} + 4 \right) }{3 \sqrt 2 |x| \left( |x| - 2 c^{}_{\phi} \right)} \;,\; 
\frac{1}{\sqrt 2} - \frac{\epsilon ~t^{}_{\phi}}{2 \sqrt 6}  \Big\} \;, \nonumber \\
 v^{}_{2} &\simeq& \Big\{ 
\frac{\epsilon}{6} \left[\frac{|x|^2-2 {\rm i} |x| s^{}_{\phi}-4}{|x|^2-2 |x| c^{}_{\phi}}+\frac{{\rm i} \sqrt{3} (|x|+2 {\rm i} s^{}_{\phi})}{|x|+2 c^{}_{\phi}}-2\right],
\nonumber\\
&& 1, \frac{\epsilon}{6} \left[\frac{|x|^2-2 {\rm i} |x| s^{}_{\phi}-4}{|x|^2-2 |x| c^{}_\phi }+\frac{\sqrt{3} (2 s^{}_{\phi}-{\rm i} |x|)}{|x|+2 c^{}_{\phi}}-2\right] \Big\}  \;,\nonumber \\
 v^{}_{3} &\simeq& \Big\{ \frac{-1}{\sqrt 2} + \frac{\epsilon  ( s^{}_\phi  + 2 {\rm i} |x| )}{2 c^{}_\phi \sqrt{6} } \;,\;
-\frac{\epsilon (2 s^{}_\phi + {\rm i} |x| )}{ \sqrt{6}	(|x| +2 c^{}_{\phi} )} \;,\; 
\frac{1}{\sqrt 2} + \frac{\epsilon~t^{}_{\phi} }{2\sqrt 6 } \Big\} \;,
\end{eqnarray}
where $s^{}_{\phi} = \sin \phi$ and $c^{}_{\phi} = \cos \phi$. The transformation matrix is $O^{}_{\nu} = \left( v^{}_{1} \;,\; v^{}_{2} \;,\; v^{}_{3} \right)^T P''$, 
where $P''$ is a diagonal phase matrix which is a part to the Majorana phase matrix in the final PMNS matrix. 
To the first order of $\epsilon$, the elements of PMNS matrix $ U^{}_{\rm PMNS} = O^{\dagger}_{l} O^{}_{\nu}$ are given by
\begin{eqnarray}
&& U^{}_{11} \simeq \sqrt{ \frac{2}{3} } + \frac{\epsilon}{3 \sqrt 2} 
\Big[ \frac{|x|^2-4 |x| c^{}_{\phi }-2{\rm i}|x| s_{\phi }+4}{\sqrt{3} |x| \left(|x|-2 c_{\phi }\right)}
- \frac{{\rm i} |x| }{c^{}_{\phi}} \Big] \;, \nonumber \\
&& U^{}_{12} \simeq \frac{1}{\sqrt{3}}-\frac{\epsilon  \left(|x|^2-4 |x| c_{\phi }^{} +2{\rm i}|x| s_{\phi }+4\right)}{3 \sqrt{3} |x| \left(|x|-2 c_{\phi}\right)} \;, \nonumber \\
&& U^{}_{13} \simeq \frac{ |x| \epsilon \left( |x| + e^{- {\rm i} \phi } \right)}{3 \sqrt{2} c_{\phi } (|x| + 2 c_{\phi })}  \;, \nonumber \\
&& U^{}_{21} \simeq \frac{-1}{\sqrt 6} + \frac{\epsilon \left[ -2 \sqrt{3} |x| \left(4 c_{\phi }+5{\rm i}s_{\phi }\right)+{\rm i}  |x|^2 \left(3 \sqrt{3} t_{\phi }+4{\rm i}\sqrt{3}-6\right)+3
	\left(\sqrt{3}+{\rm i} \right) |x|^3 \sec (\phi )+8 \sqrt{3} \right] }{18 \sqrt 2 |x| \left( |x| - 2 c^{}_{\phi} \right)}
\;, \nonumber \\
&& U^{}_{22} \simeq \frac{1}{\sqrt 3} - \frac{\epsilon \left[ |x|^3 -2 |x|^2 e^{-{\rm i} \phi} + 2 |x| e^{-2 {\rm i} \phi} -4 c_{\phi }\right]}
{ \sqrt 3 \left( |x|^3 - 4 |x| c^2_{\phi} \right)} 
\;, \nonumber \\
&& U^{}_{23} \simeq \frac{-1}{\sqrt 2} + \frac{{\rm i} \epsilon \left[ 6{\rm i}s_{\phi }+{\rm i}  |x| \left(t_{\phi }+2 \sqrt{3}+4 {\rm i} \right)
	+ \left(\sqrt{3} {\rm i} - 1\right) |x|^2 \sec (\phi ) \right] }{6 \sqrt{2} (|x| +2 c_{\phi }^{} )} 
\;, \nonumber \\
&& U^{}_{31} \simeq \frac{-1}{\sqrt 6} + \frac{ \epsilon \left[ 2{\rm i}\sqrt{3} |x| \left(s_{\phi }+4{\rm i}c_{\phi }\right)+|x|^2 \left(-3{\rm i}\sqrt{3} t_{\phi }+8 \sqrt{3}-6 {\rm i} \right)-3	\left(\sqrt{3}-{\rm i} \right) |x|^3 \sec (\phi )+8 \sqrt{3} \right] }
{18 \sqrt 2 |x| \left( |x| - 2 c^{}_{\phi} \right)}
\;, \nonumber \\
&& U^{}_{32} \simeq \frac{1}{\sqrt 3} + \frac{2 \epsilon \left( -2 |x|^2 e^{-{\rm i}\phi} + 2 c_{\phi } + |x|^3 - |x| e^{2{\rm i}\phi} \right) }{ 3\sqrt {3} \left(|x|^3-4 |x| c_{\phi }^2\right) } 
\;, \nonumber \\
&& U^{}_{33} \simeq \frac{1}{\sqrt 2} + \frac{\epsilon \left[ 6{\rm i}s_{\phi }+{\rm i}  |x| \left(t_{\phi }-2 \sqrt{3}+4 {\rm i} \right)+\left(-1-{\rm i}  \sqrt{3}\right) |x|^2 \sec (\phi ) \right]}
{6 \sqrt{2} \left( |x| + 2 c^{}_{\phi} \right) }
\;.
\end{eqnarray}
The PMNS matrix return to Eq. (\ref{TB}) up to some diagonal phase matrices when $\epsilon$ vanish. One can obtain the three mixing angles and the CP-violating phase as 
\begin{eqnarray}
s^{ }_{13} &\simeq& \frac{|x|  \epsilon \sqrt{ 2 |x| c_{\phi }+|x|^2+1 }}{3 \sqrt{2} ~c_{\phi } \left(2 c_{\phi }+|x|\right)} \;, \nonumber \\
t^{}_{23} &\simeq& 1 +  \frac{\epsilon |x| \left(4 c_{\phi } + |x| \right)}{3 c_{\phi } \left(2 c_{\phi }+|x|\right)} \;, \nonumber \\
t^{}_{12} &\simeq& \frac{1}{\sqrt{2}}-\frac{\epsilon  \left(-4 |x| c_{\phi }+|x|^2+4\right)}{2 \sqrt{2}~ |x| \left(|x|-2 c_{\phi }\right)} \;, \nonumber \\
s^{}_{\delta} &\simeq&  \sqrt{ \frac{ s^{}_{\phi}}{|x|^2 + 2 |x| c^{}_{\phi} + 1 } }
\Big\{ 1 - \frac{ \epsilon  \left( |x| c^{}_{\phi} + 1 \right) \left[ \sqrt 3 |x| \left( |x| - 2 c^{}_{\phi} \right) + 6 c^{}_{\phi} s^{}_{\phi} \right] }
{ 6 s^{}_{\phi} c^{}_{\phi} \left( |x| - 2 c^{}_{\phi} \right) } \Big\} \;. \label{83}
\end{eqnarray}
The $3~\sigma$ allowed range of mass squared difference are~\cite{Esteban:2018azc}
\begin{eqnarray}
\Delta m^2_{21} = \left( 6.79-8.01 \right) \times 10^{-5} {\rm eV}^2 \;,\qquad \Delta m^2_{31} = \left( 2.431\to 2.622 \right) \times 10^{-3} {\rm eV}^2 \;,
\end{eqnarray}
in normal mass ordering (NO) and 
\begin{eqnarray}
\Delta m^2_{21} = \left( 6.79-8.01 \right) \times 10^{-5} {\rm eV}^2 \;,\qquad \Delta m^2_{32} = \left( -2.606\to 2.413 \right)  \times 10^{-3} {\rm eV}^2 \;,
\end{eqnarray}
in inverted mass ordering (IO). By comparing the above data to the eigenvalues in Eq. (\ref{80}), we can obtain the constraints of $x$ and $\phi$, as shown in figure 3.
The lightest neutrino mass can be constrained as $0.004~{\rm eV} \leq m^{}_{1} \leq 0.006$ eV in the NO case and $0.02~{\rm eV} \leq m^{}_{3} \leq0.10$ eV in the IO case.
To illustrate the RGE effects of the three mixing angles, we define the following rescaled deviations 
\begin{eqnarray}
\Delta^{}_{13} = \frac{s^{ }_{13}}{\epsilon} \;,\quad 
\Delta^{}_{23} = \frac{t^{ }_{23}-1}{\epsilon} \;,\quad 
\Delta^{}_{12} = \frac{t^{ }_{12}-1/\sqrt{2}}{\epsilon} \;.
\end{eqnarray}
In figure 4, we present $\Delta^{}_{13}$, $\Delta^{}_{23}$ and $\Delta^{}_{12}$ as functions of the lightest neutrino mass.
The results show that $\Delta^{}_{12}$ in the IO case can reach 247.8, this is because that $x$ approximate $2 c^{}_{\phi}$ in the IO case and $\Delta^{}_{12}$ is proportional to $\left(x-2 c^{}_{\phi}\right)^{-1}$ as shown in Eq. (\ref{83}).
The other deviations are of order $\mathcal{O}(0.1)$ or $\mathcal{O}(1)$. Since the maximal value of $\epsilon$ is about $-0.004$ when $\Lambda = 10^{14}$~GeV in the MSSM,
then $\theta^{}_{23}$ and $\theta^{}_{12}$ in this model are stable against radiative corrections.

\begin{figure}[t]
	\begin{center}
		\includegraphics[scale=0.5]{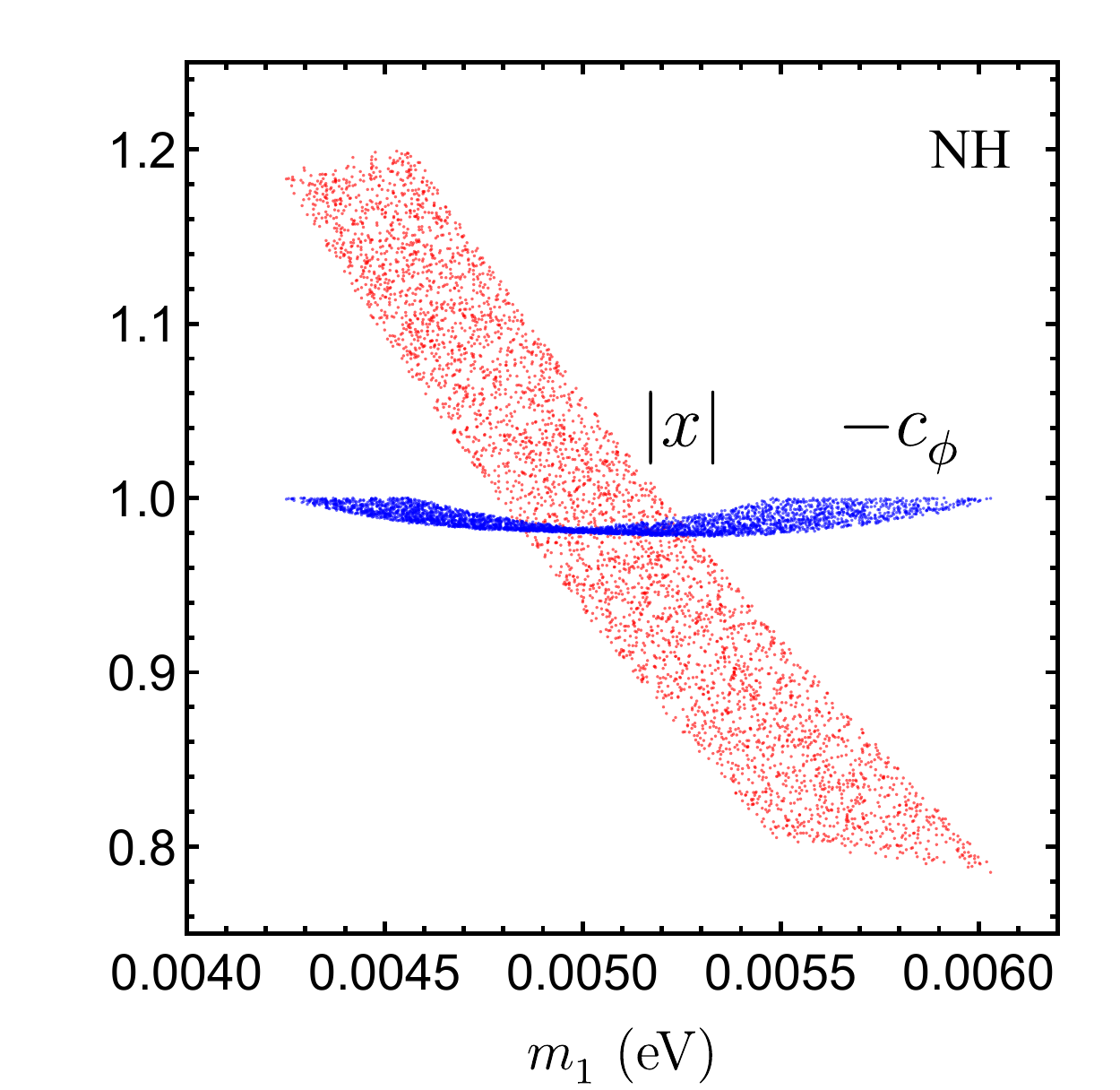} \hspace{1.cm}	
		\includegraphics[scale=0.5]{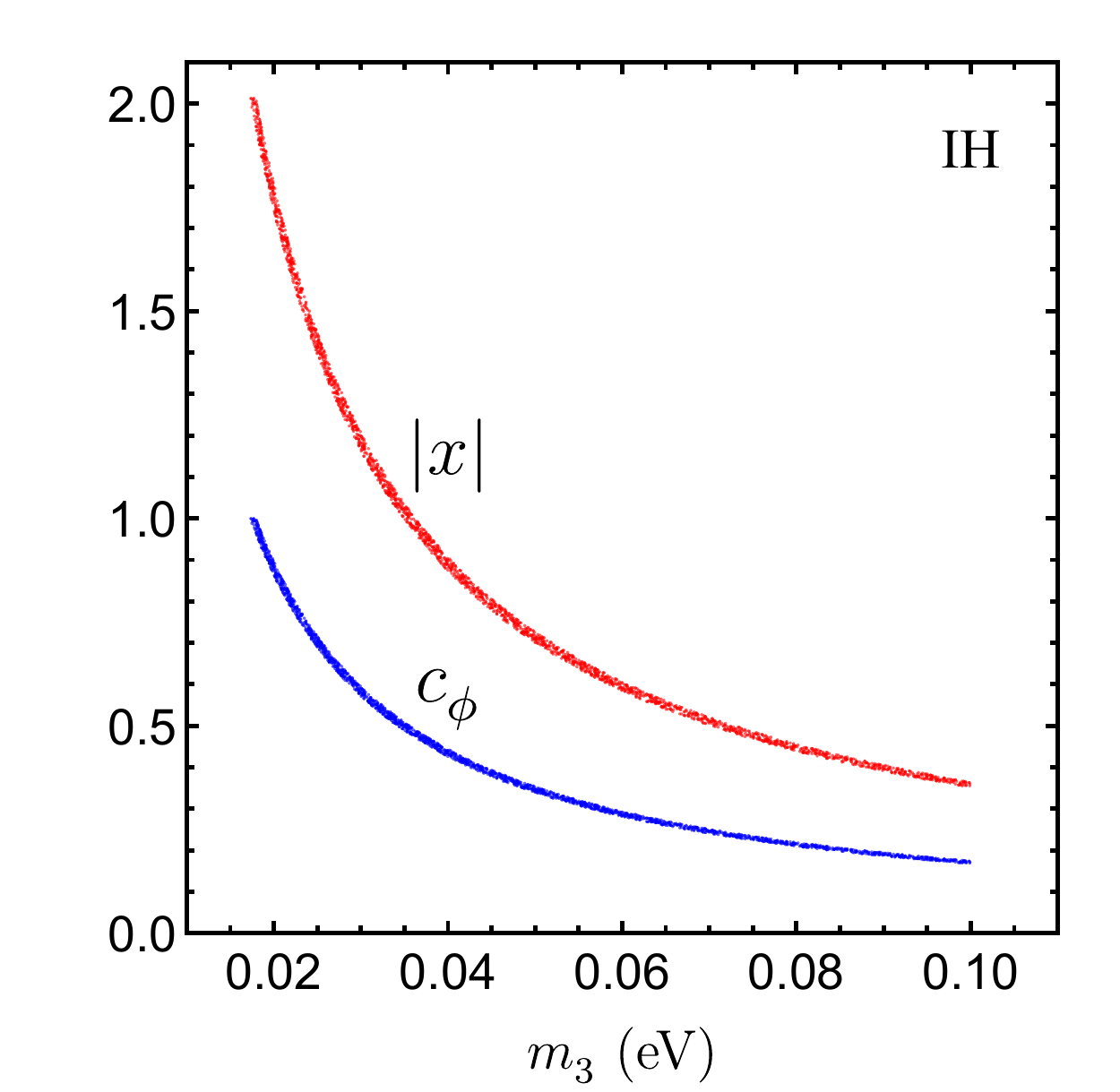} 	
		\caption{The $3~\sigma$ allowed range of of $x$ and $\phi$ when $\Lambda = 10^{14}$~GeV in the MSSM. } 
	\end{center}
\label{fig3}
\end{figure}
\begin{figure}[t]
	\begin{center}
		\includegraphics[scale=0.5]{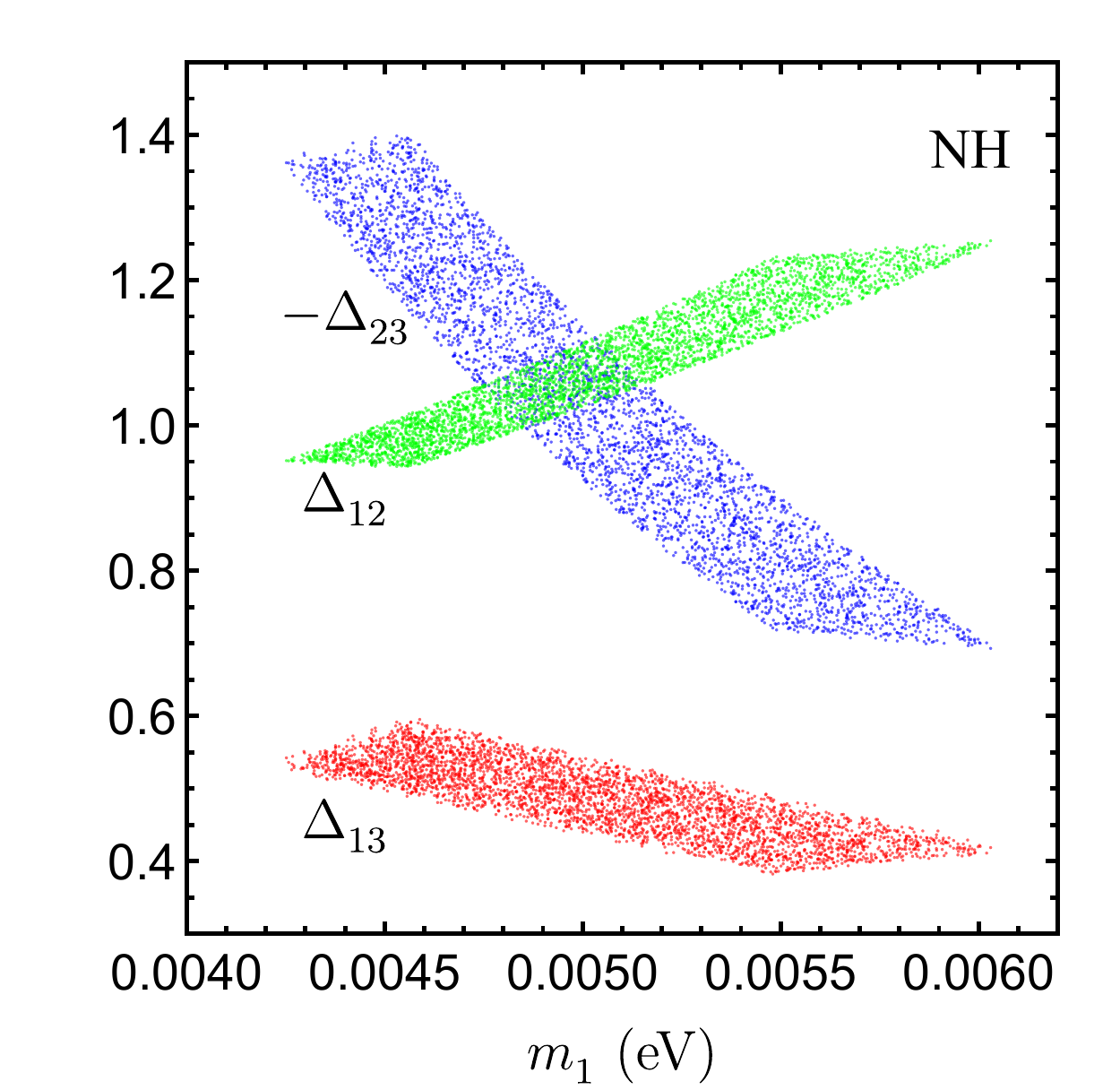} \hspace{1.cm}	
		\includegraphics[scale=0.5]{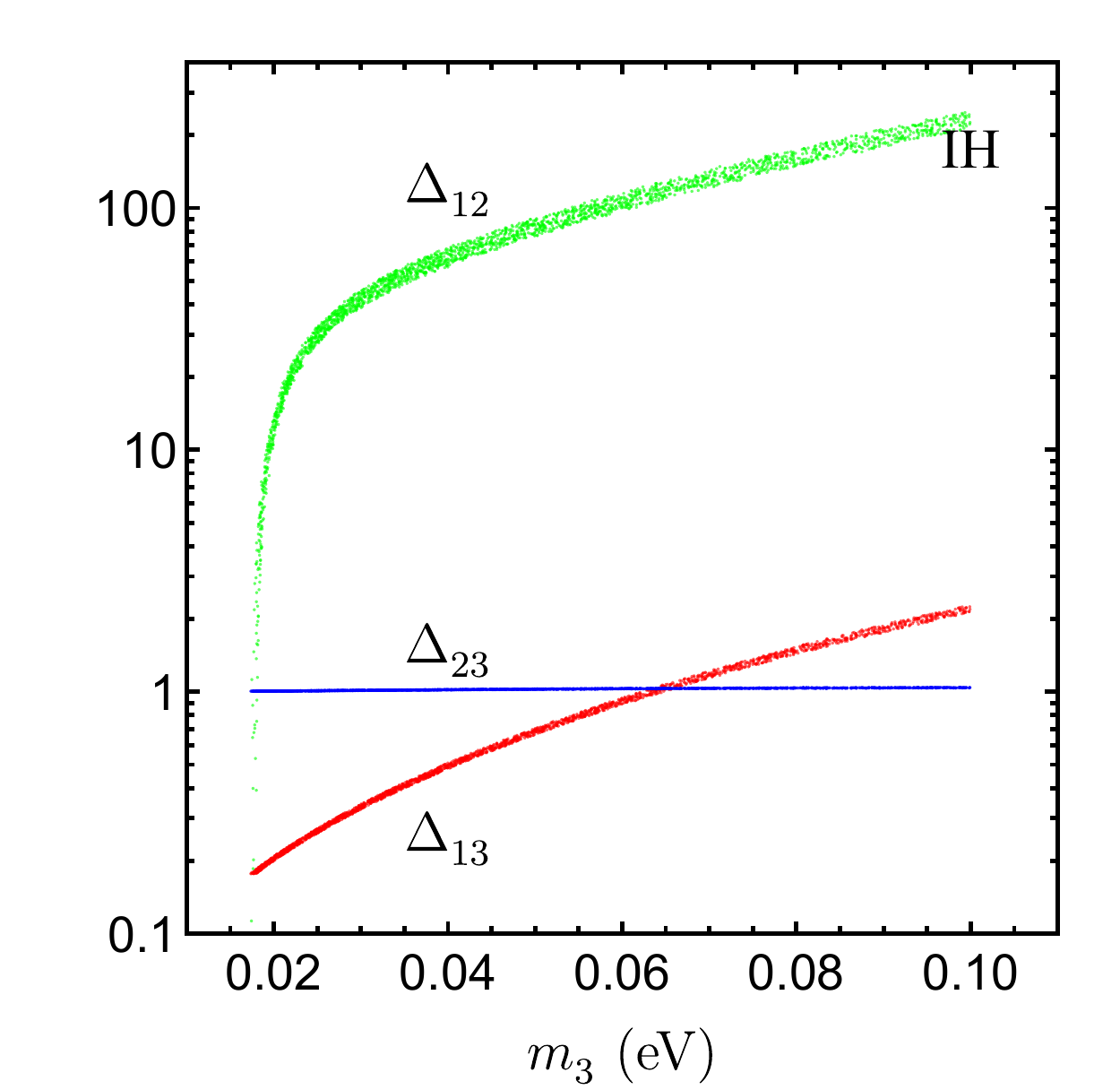} 	
		\caption{The rescaled deviations of three mixing angles as functions of the lightest neutrino mass. } 
	\end{center}
	\label{fig4}
\end{figure}

\section{Summary}
We have studied one-loop renormalization-group equations (RGEs) evolution of quark and lepton mass matrices with general structures simultaneously, where both Dirac and Majorana neutrinos are considered. Provided that the non-linear terms of RGEs are dominated by the Yukawa couplings of top quark and $\tau$ lepton, 
the unitary matrices $O^{}_{{\rm u},l}$ that diagonalize the mass matrices of up-type quarks and charged leptons in Hermitian basis are found as invariant with energy scale.
Based on this result, we can decouple the RGEs and obtain the RGE corrected masses matrices of fermion. $I^{}_{\rm f}$ (for ${\rm f = u,d,}l,\nu$) and $O^{}_{{\rm u},l}$ can modify the structures of fermion mass matrices, while only $O^{}_{{\rm u},l}$, or more specifically the third column of $O^{}_{{\rm u},l}$ , may effect the texture zeros of fermion mass matrices.
When $M^{}_{{\rm u},l}\left(t^{}_{0}\right) $ (i.e. $O^{}_{{\rm u},l}$) have some simple structures, the possible texture zeros of some fermion mass matrices
could be invariant with scale in a reliable approximation. Such as the quark mass matrices with four texture zeros, parts of the RRR pattern with five texture zeros 
and the neutrino mass matrix in FTY model. Finally, we consider the RGE running effect of the tribimaximal mixing pattern, the result show that only $\theta^{}_{12}$ might obtain 
a significant deviation.

\vspace{0.5cm}
I would like to thank Prof. Zhi-zhong Xing for providing this topic and to thank Prof. Shun Zhou for reading the manuscript.
I am also grateful to Di Zhang, Guo-yuan Huang and Jing-yu Zhu for helpful discussions.
This work was supported in part by the National Natural Science Foundation of China under grant No. 11775231
and grant No. 11835013.


\newpage
\begin{thebibliography}{99}

%\cite{Albright:1988im} 
\bibitem{Albright:1988im}
C.~H.~Albright and M.~Lindner,
%``Nonlinear Evolution of the Three Family Fritzsch Mass Matrices,''
Phys.\ Lett.\ B {\bf 213} (1988) 347.
%doi:10.1016/0370-2693(88)91773-X
%%CITATION = doi:10.1016/0370-2693(88)91773-X;%%
%15 citations counted in INSPIRE as of 23 May 2019

%\cite{Grzadkowski:1987tf}
\bibitem{Grzadkowski:1987tf}
B.~Grzadkowski and M.~Lindner,
%``Nonlinear Evolution of Yukawa Couplings,''
Phys.\ Lett.\ B {\bf 193} (1987) 71.
%doi:10.1016/0370-2693(87)90458-8
%%CITATION = doi:10.1016/0370-2693(87)90458-8;%%
%57 citations counted in INSPIRE as of 23 May 2019

%\cite{Grzadkowski:1987wr}
\bibitem{Grzadkowski:1987wr}
B.~Grzadkowski, M.~Lindner and S.~Theisen,
%``Nonlinear Evolution of Yukawa Couplings in the Double Higgs and Supersymmetric Extensions of the Standard Model,''
Phys.\ Lett.\ B {\bf 198} (1987) 64.
%doi:10.1016/0370-2693(87)90160-2
%%CITATION = doi:10.1016/0370-2693(87)90160-2;%%
%72 citations counted in INSPIRE as of 23 May 2019

%\cite{Giudice:1992an}
\bibitem{Giudice:1992an}
G.~F.~Giudice,
%``A New ansatz for quark and lepton mass matrices,''
Mod.\ Phys.\ Lett.\ A {\bf 7} (1992) 2429.
%doi:10.1142/S0217732392003876
%[hep-ph/9204215].
%%CITATION = doi:10.1142/S0217732392003876;%%
%80 citations counted in INSPIRE as of 23 May 2019

%\cite{Xing:2015sva}
\bibitem{Xing:2015sva}
Z.~z.~Xing and Z.~h.~Zhao,
%``On the four-zero texture of quark mass matrices and its stability,''
Nucl.\ Phys.\ B {\bf 897} (2015) 302.
%doi:10.1016/j.nuclphysb.2015.05.027
%[arXiv:1501.06346 [hep-ph]].
%%CITATION = doi:10.1016/j.nuclphysb.2015.05.027;%%
%13 citations counted in INSPIRE as of 22 May 2019

%\cite{Lindner:2005as}
\bibitem{Lindner:2005as} 
M.~Lindner, M.~Ratz and M.~A.~Schmidt,
%``Renormalization group evolution of Dirac neutrino masses,''
JHEP {\bf 0509}, 081 (2005).
%doi:10.1088/1126-6708/2005/09/081
%[hep-ph/0506280].
%%CITATION = doi:10.1088/1126-6708/2005/09/081;%%
%35 citations counted in INSPIRE as of 10 Jun 2019

%\cite{Xing:2011zza}
\bibitem{Xing:2011zza} 
Z. z. Xing and S. Zhou,
%``Neutrinos in particle physics, astronomy and cosmology,''
Springer-Verlag, Berlin Heidelberg (2011).
%39 citations counted in INSPIRE as of 17 Apr 2019

%\cite{Xing:2011aa}
\bibitem{Xing:2011aa} 
Z.~z.~Xing, H.~Zhang and S.~Zhou,
%``Impacts of the Higgs mass on vacuum stability, running fermion masses and two-body Higgs decays,''
Phys.\ Rev.\ D {\bf 86}, 013013 (2012).
%doi:10.1103/PhysRevD.86.013013
%[arXiv:1112.3112 [hep-ph]].
%%CITATION = doi:10.1103/PhysRevD.86.013013;%%
%128 citations counted in INSPIRE as of 11 Jun 2019

%\cite{Tanabashi:2018oca}
\bibitem{Tanabashi:2018oca} 
M.~Tanabashi {\it et al.} [Particle Data Group],
%``Review of Particle Physics,''
Phys.\ Rev.\ D {\bf 98}, no. 3, 030001 (2018).
%doi:10.1103/PhysRevD.98.030001
%%CITATION = doi:10.1103/PhysRevD.98.030001;%%
%1794 citations counted in INSPIRE as of 11 Jun 2019

%\cite{Xing:1996hi}
\bibitem{Xing:1996hi}
Z.~z.~Xing,
%``Implications of the quark mass hierarchy on flavor mixings,''
J.\ Phys.\ G {\bf 23} (1997) 1563.
%doi:10.1088/0954-3899/23/11/006
%[hep-ph/9609204].
%%CITATION = doi:10.1088/0954-3899/23/11/006;%%
%45 citations counted in INSPIRE as of 26 Aug 2019


%\cite{Fritzsch:1979zq}
\bibitem{Fritzsch:1979zq} 
H. Fritzsch,
%``Quark Masses and Flavor Mixing,''
Nucl.\ Phys.\ B {\bf 155} (1979) 189.
%doi:10.1016/0550-3213(79)90362-6
%%CITATION = doi:10.1016/0550-3213(79)90362-6;%%
%698 citations counted in INSPIRE as of 24 Apr 2019

%\cite{Fritzsch:1999rb}
\bibitem{Fritzsch:1999rb} 
H. Fritzsch and Z. z. Xing,
%``The Light quark sector, CP violation, and the unitarity triangle,''
Nucl.\ Phys.\ B {\bf 556} (1999) 49.
%doi:10.1016/S0550-3213(99)00337-5
%[hep-ph/9904286].
%%CITATION = doi:10.1016/S0550-3213(99)00337-5;%%
%111 citations counted in INSPIRE as of 24 Apr 2019

%\cite{Branco:1999nb}
\bibitem{Branco:1999nb} 
G. C. Branco, D. Emmanuel-Costa and R. Gonzalez Felipe,
%``Texture zeros and weak basis transformations,''
Phys.\ Lett.\ B {\bf 477} (2000) 147.
%doi:10.1016/S0370-2693(00)00193-3
%[hep-ph/9911418].
%%CITATION = doi:10.1016/S0370-2693(00)00193-3;%%
%91 citations counted in INSPIRE as of 24 Apr 2019

%\cite{Fritzsch:1977vd}
\bibitem{Fritzsch:1977vd}
H.~Fritzsch,
%``Weak Interaction Mixing in the Six - Quark Theory,''
Phys.\ Lett.\  {\bf 73B} (1978) 317;
%doi:10.1016/0370-2693(78)90524-5
%%CITATION = doi:10.1016/0370-2693(78)90524-5;%%
%836 citations counted in INSPIRE as of 22 May 2019
%\cite{Fritzsch:1979zq}
%\bibitem{Fritzsch:1979zq}
%H.~Fritzsch,
%``Quark Masses and Flavor Mixing,''
Nucl.\ Phys.\ B {\bf 155} (1979) 189.
%doi:10.1016/0550-3213(79)90362-6
%%CITATION = doi:10.1016/0550-3213(79)90362-6;%%
%698 citations counted in INSPIRE as of 22 May 2019

%\cite{Fukugita:1992sy}
\bibitem{Fukugita:1992sy}
M.~Fukugita, M.~Tanimoto and T.~Yanagida,
%``Phenomenological lepton mass matrix,''
Prog.\ Theor.\ Phys.\  {\bf 89} (1993) 263.
%doi:10.1143/PTP.89.263
%%CITATION = doi:10.1143/PTP.89.263;%%
%63 citations counted in INSPIRE as of 22 May 2019

%\cite{Fukugita:2003tn}
\bibitem{Fukugita:2003tn}
M.~Fukugita, M.~Tanimoto and T.~Yanagida,
%``Predictions from the Fritzsch type lepton mass matrices,''
Phys.\ Lett.\ B {\bf 562} (2003) 273;
%doi:10.1016/S0370-2693(03)00568-9
%[hep-ph/0303177].
%%CITATION = doi:10.1016/S0370-2693(03)00568-9;%%
%50 citations counted in INSPIRE as of 22 May 2019
%\cite{Fukugita:2012jr}
%\bibitem{Fukugita:2012jr}
M.~Fukugita, Y.~Shimizu, M.~Tanimoto and T.~T.~Yanagida,
%``θ$_{13}$ in neutrino mass matrix with the minimal texture,''
Phys.\ Lett.\ B {\bf 716} (2012) 294;
%doi:10.1016/j.physletb.2012.06.049
%[arXiv:1204.2389 [hep-ph]].
%%CITATION = doi:10.1016/j.physletb.2012.06.049;%%
%28 citations counted in INSPIRE as of 22 May 2019
%\cite{Xing:2004xu}
%\bibitem{Xing:2004xu}
Z.~z.~Xing and S.~Zhou,
%``Isomeric lepton mass matrices and bilarge neutrino mixing,''
Phys.\ Lett.\ B {\bf 593} (2004) 156;
%doi:10.1016/j.physletb.2004.04.059
%[hep-ph/0403261].
%%CITATION = doi:10.1016/j.physletb.2004.04.059;%%
%29 citations counted in INSPIRE as of 23 May 2019
%\cite{Zhou:2004wz}
%\bibitem{Zhou:2004wz}
S.~Zhou and Z.~z.~Xing,
%``A Systematic study of neutrino mixing and CP violation from lepton mass matrices with six texture zeros,''
Eur.\ Phys.\ J.\ C {\bf 38} (2005) 495.
%doi:10.1140/epjc/s2004-02065-2
%[hep-ph/0404188].
%%CITATION = doi:10.1140/epjc/s2004-02065-2;%%
%51 citations counted in INSPIRE as of 23 May 2019

%\cite{Xing:2014sja}
\bibitem{Xing:2014sja}
Z.~z.~Xing,
%``Quark Mass Hierarchy and Flavor Mixing Puzzles,''
Int.\ J.\ Mod.\ Phys.\ A {\bf 29} (2014) 1430067;
%doi:10.1142/S0217751X14300671
%[arXiv:1411.2713 [hep-ph]].
%%CITATION = doi:10.1142/S0217751X14300671;%%
%15 citations counted in INSPIRE as of 22 May 2019
N.~Mahajan, R.~Verma and M.~Gupta,
%``Investigating non-Fritzsch like texture specific quark mass matrices,''
Int.\ J.\ Mod.\ Phys.\ A {\bf 25} (2010) 2037;
W.~A.~Ponce and R.~H.~Benavides,
%``Texture Zeros for the Standard Model Quark Mass Matrices,''
Eur.\ Phys.\ J.\ C {\bf 71} (2011) 1641;
Y.~Giraldo,
%``Texture Zeros and WB Transformations in the Quark Sector of the Standard Model,''
Phys.\ Rev.\ D {\bf 86} (2012) 093021.

%\cite{Fritzsch:1986sn}
\bibitem{Fritzsch:1986sn}
H.~Fritzsch,
%``Hierarchical Chiral Symmetries and the Quark Mass Matrix,''
Phys.\ Lett.\ B {\bf 184} (1987) 391.
%doi:10.1016/0370-2693(87)90186-9
%%CITATION = doi:10.1016/0370-2693(87)90186-9;%%
%95 citations counted in INSPIRE as of 23 May 2019

%\cite{Fritzsch:1997fw}
\bibitem{Fritzsch:1997fw}
H.~Fritzsch and Z.~Z.~Xing,
%``Flavor symmetries and the description of flavor mixing,''
Phys.\ Lett.\ B {\bf 413} (1997) 396;
Phys.\ Rev.\ D {\bf 57} (1998) 594;
Phys.\ Lett.\ B {\bf 555} (2003) 63.


%\cite{Fritzsch:1999ee}
\bibitem{Fritzsch:1999ee}
H.~Fritzsch and Z.~z.~Xing,
%``Mass and flavor mixing schemes of quarks and leptons,''
Prog.\ Part.\ Nucl.\ Phys.\  {\bf 45} (2000) 1;
%doi:10.1016/S0146-6410(00)00102-2
%[hep-ph/9912358].
%%CITATION = doi:10.1016/S0146-6410(00)00102-2;%%
%502 citations counted in INSPIRE as of 22 May 2019

%\cite{Ibanez:1994ig}
\bibitem{Ibanez:1994ig}
L.~E.~Ibanez and G.~G.~Ross,
%``Fermion masses and mixing angles from gauge symmetries,''
Phys.\ Lett.\ B {\bf 332} (1994) 100;
B.~R.~Desai and A.~R.~Vaucher,
%``Quark mass matrices with four and five texture zeroes, and the CKM matrix, in terms of mass eigenvalues,''
Phys.\ Rev.\ D {\bf 63} (2001) 113001;
H.~D.~Kim, S.~Raby and L.~Schradin,
%``Quark mass textures and sin 2 beta,''
Phys.\ Rev.\ D {\bf 69} (2004) 092002;
W.~A.~Ponce, J.~D.~Gómez and R.~H.~Benavides,
%``Five texture zeros and CP violation for the standard model quark mass matrices,''
Phys.\ Rev.\ D {\bf 87} (2013) no.5,  053016.

%\cite{Ramond:1993kv}
\bibitem{Ramond:1993kv}
P.~Ramond, R.~G.~Roberts and G.~G.~Ross,
%``Stitching the Yukawa quilt,''
Nucl.\ Phys.\ B {\bf 406} (1993) 19
%doi:10.1016/0550-3213(93)90159-M
[hep-ph/9303320].
%%CITATION = doi:10.1016/0550-3213(93)90159-M;%%
%401 citations counted in INSPIRE as of 24 May 2019


%\cite{Xing:2003yj}
\bibitem{Xing:2003yj}
Z.~z.~Xing and H.~Zhang,
%``Complete parameter space of quark mass matrices with four texture zeros,''
J.\ Phys.\ G {\bf 30} (2004) 129.
%doi:10.1088/0954-3899/30/2/011
%[hep-ph/0309112].
%%CITATION = doi:10.1088/0954-3899/30/2/011;%%
%55 citations counted in INSPIRE as of 22 May 2019
	
%\cite{Babu:2005se}
\bibitem{Babu:2005se} 
K.~S.~Babu and X.~G.~He,
%``Model of geometric neutrino mixing,''
hep-ph/0507217.
%%CITATION = HEP-PH/0507217;%%
%193 citations counted in INSPIRE as of 08 Aug 2019
	
%\cite{Esteban:2018azc}
\bibitem{Esteban:2018azc}
I.~Esteban, M.~C.~Gonzalez-Garcia, A.~Hernandez-Cabezudo, M.~Maltoni and T.~Schwetz,
%``Global analysis of three-flavour neutrino oscillations: synergies and tensions in the determination of $\theta_23, \delta_CP$, and the mass ordering,''
JHEP {\bf 1901} (2019) 106.
%doi:10.1007/JHEP01(2019)106
%[arXiv:1811.05487 [hep-ph]].
%%CITATION = doi:10.1007/JHEP01(2019)106;%%
%150 citations counted in INSPIRE as of 25 Aug 2019	
	
\end{thebibliography}
\end{document}